\documentclass[reprint,
onecolumn,
superscriptaddress,
nofootinbib,
aps,
prfluids,
]{revtex4-2}

\usepackage[dvipsnames]{xcolor}
\usepackage{bm}
\usepackage{tikz}
\usetikzlibrary{arrows.meta}
\usepackage{longtable}
\usepackage{graphicx}
\usepackage{epstopdf}
\usepackage{hyperref}
\usepackage{grffile}
\usepackage{amsmath}
\usepackage{amssymb}
\usepackage{comment}
\usepackage{float}

\newcommand{\br}{{\bf r}}
\newcommand{\bu}{{\bf u}}

\newcommand{\mC}{{\mathcal C}}
\newcommand{\mD}{{\mathcal D}}
\newcommand{\mH}{{\mathcal H}}
\newcommand{\mM}{{\mathcal M}}

\newcommand{\beq}{\begin{equation}}
\newcommand{\eeq}{\end{equation}}

\newcommand{\scal}[1]{\left\langle #1\right\rangle}

\newcommand{\inphyni}{Institut de Physique de Nice, Universit\'e C\^ote d'Azur CNRS - UMR 7010, 17 rue Julien Laupr\^etre, 06200 Nice, France}
\newcommand{\birm}{Department of Mechatronics and Biomedical Engineering, College of Engineering and Physical Sciences,
Aston University, Birmingham, B4 7ET, United Kingdom}
\newcommand{\shen}{Department of Mechanics and Aerospace Engineering, and Centre for Complex Flow and Soft Matter, Southern University of Science and Technology, Shenzhen 518055, China}
\graphicspath{{figs/}}
\begin{document}

\title{Emergence of Zonal Flows in an Imbalanced Point-Vortex System}

\author{Jorge Carreño-Donoso}
\email{jorge.carreno-donoso@univ-cotedazur.fr}
\affiliation{\inphyni}
\author{Simon Thalabard}
\email{simon.thalabard@univ-cotedazur.fr}
\affiliation{\inphyni}
\author{Ying Zhu}
\affiliation{\inphyni}
\affiliation{\shen}
\author{Thomas Frisch}
\affiliation{\inphyni}
\author{Jason Laurie}
\affiliation{\birm}
\author{Sergey Nazarenko}
\email{sergey.nazarenko@univ-cotedazur.fr}
\affiliation{\inphyni}

\date{\today}

\begin{abstract}
\noindent Motivated by experiments on rotating Bose–Einstein condensates, we investigate the statistical mechanics and the non-equilibrium dynamics of an imbalanced gas of oppositely signed Onsager point vortices confined in a disk.
To model external forcing and dissipation, we supplement their Hamiltonian dynamics with an annihilation-reinjection  (AR) mechanism that preserves the imbalance: oppositely signed vortices are annihilated when they approach closer than a distance representing the healing length, and are subsequently reintroduced into the domain at random locations.
At positive energies, the system self-organizes into a  large-scale coherent dipole at low imbalance and into a zonal flow at higher imbalance. While AR breaks  energy and angular momentum conservation, we argue that the vortex gas evolves quasi-adiabatically through the equilibrium phase space. We capture  the topological transition between dipolar and zonal flows both from a statistical and a hydrodynamical perspective. In particular, the bifurcation reflects a  non-trivial equilibrium diagram  in terms of  energy $e$, angular momentum $\mu$, and vortex imbalance $\eta$. We characterize this phase diagram  in the ``strong mixing'' regime  studied by linearizing the Joyce-Montgomery  mean-field theory about the weakly non-uniform state. We  show that the linear theory is surprisingly relevant for the numerical simulations with AR and suggest that the selection between the dipolar and the  zonal configurations in the $(e,\mu,\eta)$ phase space is driven by a hydrodynamical shear instability.

\begin{description}
\item[Subject Areas]
Fluid Mechanics, Statistical Physics, Condensed Matter Physics.
\end{description}
\end{abstract}

\maketitle
\tableofcontents
\section{Introduction}
\noindent The statistical behavior of a gas of hydrodynamic point vortices confined to a bounded two-dimensional (2D) domain is a classical problem in  hydrodynamics~\cite{EyinkSreenivasan2006Onsager,pointin1976statistical,Newton2001}. It historically connects to the problem of classical 2D turbulence~\cite{GeldhauserRomito}, and also serves as a fundamental model for quantized vortices in superfluids~\cite{reeves2013onsager, ReevesBillamYuBradley2017EnstrophyCascade, YuBillamNianReevesBradley2016VortexClustering,LydonNazarenkoLaurie2022DipoleDynamicsPointVortex, Skipp2023HamiltonianPointVortex}.  Experimentally, vortex gases can be realized in trapped Bose-Einstein condensates (BEC), where quantized vortices are nucleated by external excitations or rotation~\cite{fetter2009rotating}. Recent advances in experimental techniques now enable the confinement of cold atoms in a variety of geometries 
\cite{Seo2017,Guo2020,MerlotiDubessyLongchambonPerrin20132DQuantumGasMagnetTrap,GuoGutierrezReyBadrPerrin2022ExpansionShellTrap}, including highly oblate and bubble-shaped traps realizing quasi-2D setups~\cite{Kwon2014, dubessy2025quantum}. In such systems, vortex populations need not be neutral: 
excitation protocols may introduce rotation and produce states with unequal numbers of vortices and antivortices, leading to a finite net circulation.
Understanding how this imbalance modifies the statistical organization of vortex gases is the central motivation of the present work.

Microscopically, a vortex gas is represented by $N$ point-like vortices with quantized circulations $\Gamma_i$  interacting through the pair potential $\sim -\Gamma_i\Gamma_j \ln(r_{ij})$, e.g., logarithmically diverging as $r_{ij} \to 0$. In bounded domains, these interactions lead to the clustering of  same-sign vortices into macroscopic clusters at high energies.  
Onsager first interpreted these states as negative temperature equilibria~\cite{Onsager1949, EyinkSreenivasan2006Onsager} later formalized  by the  mean-field (MF) theory of Joyce and Montgomery~\cite{JoyceMontgomery1973,EyinkSreenivasan2006Onsager}. Onsager's theory  subsequently  inspired the  statistical theory of  continuum 2D \cite{Carnevale1991Evolution} and related geophysical flows~\cite{Miller1990,Robert1990,Robert1991,Robert2000,RobertSommeria1991,turkington1999statistical,bouchet2012statistical}.
Previous numerical studies have examined the role of the angular momentum~\cite{pointin1976statistical,ReevesGoddardLeeGauthier2022TurbulentRelaxation,SmithONeil1990}, finite size effects and relaxation dynamics~\cite{yatsuyanagi2005dynamics}, and effects of geometry~\cite{MaestriniSalman2019}. 
To our knowledge, however, most studies have focused either on neutral systems {(where there are equal numbers of positively- and negatively-signed vortices)}
or on {single-signed vortex systems}, 
leaving the role of imbalance between the numbers of positive and negative vortices largely unexplored.

In the present paper,  we investigate both the equilibrium statistical mechanics and the non-equilibrium dynamics of a two-signed vortex gas confined to a disk, with a net imbalance between positive and negative vortices. All circulations have the same magnitude as is the case in superfluids. 
In the equilibrium case, 
we combine microcanonical Monte-Carlo (MC) simulations based on the Creutz algorithm with numerical analysis of the Joyce-Montgomery theory to characterize the large-scale equilibrium topologies at prescribed energy and angular momentum.   Beyond these, %

our central control parameter is the imbalance parameter 
\begin{equation}
	\eta = \dfrac{N^{+}-N^{-}}{N} \in [0,1],
	\label{eq:imbalance}
\end{equation}
where $N^+$ and $N^-$ are the numbers of positively- and negatively-signed vortices respectively, assuming, without loss of generality, that $N^+\geq N^-$, and $N=N^++N^-$ being the total number of vortices.
From the numerical  analysis of the Joyce-Montgomery theory  in the  quasi-uniform, high-temperature regime,  we show how increasing $\eta$ from zero to one alters the  phase diagram classifying  the various flow topologies as a function of the energy and angular momentum, and may drive transitions  from dipolar toward zonal-flow configurations---reminiscent of large-scale jets observed in planetary atmospheres~\cite{galperin2019zonal}.

Beyond equilibrium statistical mechanics, we also investigate non-equilibrium dynamics motivated by Bose-Einstein condensate experiments. Unlike classical point-vortex dynamics, which is Hamiltonian and conserves both energy and circulation, quantum vortex gases exhibit dissipation through vortex--antivortex annihilation as seen in both experiments~\cite{Seo2017, gauthier2019giant, Kwon2014} and  numerical simulations~\cite{NazarenkoOnorato2006WaveTurbulenceBEC,Nazarenko2006,simula2014emergence,StaggAllenParkerBarenghi2015QTTBEC,gauthier2019giant}.
Each annihilation removes two vortices while converting the negative energy associated with the hollow vortex cores into sound, thereby increasing the average energy per remaining vortex. This mechanism is known as evaporative heating~\cite{simula2014emergence}.
To mimic this process, we augment the Hamiltonian point-vortex dynamics with the annihilation--reinjection (AR) mechanism in the spirit of the evaporative heating modeled in \cite{MaestriniSalman2019}. Vortex--antivortex pairs whose separation falls below a threshold distance $\varepsilon$ are removed, while two new vortices, positive and negative,  are reinjected at random independent positions. This mechanism generates fluctuations in both the energy and angular momentum and, in principle, drives the system into statistically stationary non-equilibrium states. 
Throughout its evolution, we find however that the dynamics remains remarkably close to the corresponding sequence of microcanonical equilibrium states, suggesting that  the point-vortex gas evolves adiabatically through the equilibrium phase space, determined by the energy $e$, the angular momentum  $\mu$ and the imbalance $\eta$. At high imbalance, the system follows a trajectory in  $(e,\mu)$ that is associated with an axisymmetric equilibrium, -- a large-scale zonal flow. At low imbalance, the system bifurcates towards values of $(e,\mu)$  associated with large-scale dipoles. We show that this bifurcation  is largely independent of the annihilation length, and discuss  an interpretation  in terms of a hydrodynamical shear instability. 

The paper is structured as follows: \S\ref{sec:formulation} recalls the Hamiltonian formulation of point vortices and presents the AR mechanism. \S\ref{sec:equilibrium} recalls the Onsager and Joyce-Montgomery equilibrium theories and describes the MF predictions for the emergence of zonal flows. \S\ref{sec:dns} discusses the numerical simulations of point vortices evolving under AR. \S\ref{sec:conclusion} formulates concluding remarks.

\section{Two-sign point-vortex system in a disk}
\label{sec:formulation}

\noindent This section recalls the Hamiltonian formulation of a point-vortex system, and describes the AR mechanism employed to drive the gas out of equilibrium.

\subsection{Hamiltonian description}
\label{sec:hamiltonian}

\noindent An assembly of $N$ hydrodynamical point vortices evolving in a disk $\mathcal{D} \subset \mathbb{R}^2$ of radius $R=1$ is driven by the Hamiltonian \cite{Newton2001}
\begin{equation}
\mathcal{H} = -\frac{1}{4\pi} {\sum^N_{\substack{j, k =1\\j\neq k}}} \Gamma_j \Gamma_k \ln |\br_j - \br_k| + \frac{1}{4\pi} \sum_{j,k=1}^N \Gamma_j \Gamma_k \ln \left| \br_j - \tilde{\br}_k \right|+\frac{1}{4\pi} \sum_{j,k=1}^N \Gamma_j \Gamma_k \ln \left| \br_k\right|.
\label{eq:hamiltonian}
\end{equation}
Here, $\br_k =(x_k, y_k)$ denotes the position of the $k$-th vortex and $\Gamma_k$ its circulation.
The first term in Eq.~\eqref{eq:hamiltonian} represents the logarithmic (Coulomb-like) interaction between vortices, while the second and third terms arise from the boundary condition and account for interactions with image vortices located at $\tilde{\br }_k = \br_k / |\br_k|^2$ \cite{Newton2001}.  The sum over distinct pairs of vortices in the first sum removes the formally constant, but infinite, vortex self-interaction contribution. Subsequently, the equation of motion  can be written in Hamiltonian form as  \cite{Helmholtz1858,Wiegmann2013HydrodynamicsEulerFQHE,Kirchhoff1883Vorlesungen,patil2021chiral}
\begin{equation}
\dot{\br }_k = -\dfrac{1}{\Gamma_k} \nabla^\perp_k \mathcal{H}=\frac{1}{2\pi} \sum^N_{\substack{j=1\\j\neq k}} \Gamma_j \frac{(\br_k-\br_j)^\perp}{|\br_k- \br_j|^2}
-\frac{1}{2\pi} \sum_{j=1}^N \Gamma_j \frac{(\br_k - \tilde \br_j)^\perp}{|\br_k - \tilde \br_j|^2}
,\quad\text{for}\quad 1\leq k\leq N,
\label{eq:dynamical_equations}
\end{equation}
where we write $\br^\perp_k := (-y_k,x_k)$ and $\nabla^\perp_k = (-\partial_{y_k},\partial_{x_k})$.
The dynamics \eqref{eq:dynamical_equations} conserve the Hamiltonian~\eqref{eq:hamiltonian} whose meaning, as usual, is the total energy. Because the disk is rotationally symmetric, the Hamiltonian is invariant under rotation
implying conservation of the angular momentum, 
\begin{equation}\label{angular}
\mathcal{M} = \sum_{k=1}^{N} \Gamma_k\, |\br_k|^2.
\end{equation}

We hereafter consider a two-signed point-vortex system consisting of $N = N^{+} + N^{-}$ vortices, where $N^{+}$ and $N^{-}$ denote the numbers of vortices with positive and negative circulations, respectively, with $\Gamma_k = \pm 1$. 
We use the imbalance parameter of Eq.~\eqref{eq:imbalance}, $\eta = (N^{+}-N^{-})/N$, which characterizes the excess of positive vortices. The parameter satisfies $0 \le \eta \le 1$, where $\eta = 0$ corresponds to a neutral vortex system, while $\eta = 1$ corresponds to a single-signed vortex system.

The Hamiltonian formulation~\eqref{eq:dynamical_equations} stems from the  Biot-Savart law applied to the singular vorticity distribution field
\begin{equation}\label{eq:om}
	\omega(\br,t) = \sum_{k=1}^{N} \Gamma_k\,\delta\left(\br- \br_k(t)\right).
\end{equation}
The corresponding 
streamfunction is 
\begin{equation}
	\psi(\br,t) = 
\sum_{k=1}^{N} \Gamma_k\,G \left(\br, \br_k(t)\right),
\end{equation}
involving the zero-boundary Green-function $G(\br,\br_k):=-\Delta^{-1}\delta(\br-\br_k)$ defined on the unit disk:
\begin{equation}
\quad G(\br,\br_k)
:=
-\frac{1}{2\pi}
\left(
\ln\left| \dfrac{\br-\br_k}{\br-\tilde \br_k}\right|
-\ln|\br_k|
\right),
\quad
\text{with}
\quad
\tilde \br_k=\frac{\br_k}{|\br_k|^2}.
\label{eq:green}
\end{equation}
The velocity field $\bu = -\nabla^\perp \psi$ 
satisfies free-slip boundary conditions at $|\br|=1$. The  point-vortex equations~\eqref{eq:dynamical_equations} follow by using the regular part of the streamfunction $\psi_{reg}:= \lim_{\br\to\br_k} \left(\psi+(\Gamma_k/2\pi) \ln | \br-\br_k| \right)$, i.e. by removing the (singular) self-interaction of the $k$-vortex. Rigorously, such a point vortex model was derived from the 2D Gross-Pitaevskii equation  in \cite{Skipp2023HamiltonianPointVortex}. This derivation assumes that the vortices are separated by distances much greater than the vortex core sizes. The streamfunction regularization in this system comes naturally as the vortex self-energies are (large but) finite due to the finiteness of the cores. 

\subsection{Annihilation and reinjection dynamics}
\label{subsec:FDS}
\noindent
In quantum BEC fluids, vortex–antivortex annihilation is an important dissipative mechanism, as seen in experiments \cite{Seo2017,gauthier2019giant,Kwon2014}
and numerical simulations based on the 2D Gross-Pitaevskii equation~\cite{NazarenkoOnorato2006WaveTurbulenceBEC,Nazarenko2006,simula2014emergence,StaggAllenParkerBarenghi2015QTTBEC,gauthier2019giant}.
In addition to decreasing the number of vortices by two in each event, annihilations 
lead to changes in the energy and angular momentum of the full system. In particular, the release of the negative potential energy of the hollow quantum vortex cores into an acoustic radiation to infinity (or boundaries) leads to a net increase of the kinetic energy stored in the vortex system. This is  the  so-called ``evaporative heating" effect~\cite{simula2014emergence}. In superfluid BECs, the annihilation distance is of the order of the healing length of the condensate, which sets the (hollow) vortex core size~\cite{MaestriniSalman2019, NazarenkoOnorato2006WaveTurbulenceBEC}. Besides, for the flow to display hydrodynamic properties, the vortex cores have to be significantly smaller than the mean inter-vortex spacing~\cite{Skipp2023HamiltonianPointVortex}. 

\begin{figure}[htp]
    \centering
    \includegraphics[scale=0.98]{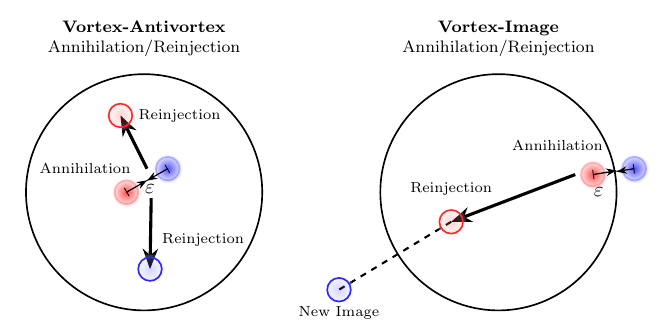}
    \caption{Sketch of the annihilation and reinjection mechanism: 
    (Left) annihilation of a vortex dipole of size $\varepsilon$ within the disk and the subsequent random reinjection of the point vortices. (Right) Annihilation of a point vortex with its image at the disk boundary and subsequent reinjection.
}
    \label{fig:anni}
\end{figure}

To model the effect of annihilations in the point-vortex system, we follow~\cite{MaestriniSalman2019}, and augment the Hamiltonian dynamics of \S\ref{sec:hamiltonian} with an ad hoc annihilation and reinjection (AR) mechanism.
Specifically,  we remove vortex–antivortex pairs when their separation falls below the annihilation distance $\varepsilon$, which plays the role of the healing length in the point-vortex setting.
Due to the presence of boundaries, a vortex may interact strongly with its image when approaching the edge of the domain. Therefore, we also include vortex–image annihilation for vortices approaching closer than $\varepsilon$  to the boundary; see both kinds of annihilation sketched in Fig.~\ref{fig:anni}.

Keeping in mind BEC applications, for the vortices to retain their hydrodynamic properties, their separations must greatly exceed the healing length. Thus we choose $\varepsilon \ll d$ where 
 $d \sim \sqrt{\pi  / N}$ is
 the mean inter-vortex spacing 
 (an estimate assuming statistically uniform distribution of vortices in the unit disk). 
The annihilation procedure decreases the number of vortices and generates fluctuations in the point-vortex energy and angular momentum. To allow for  statistically stationary states, we introduce a forcing mechanism that reinjects vortices back into the system. In real BEC experiments, forcing may take different forms, e.g., by shaking the confining trap or through moving an obstacle using an ``optical spoon" inside the domain.  In our setting, we implement a reinjection process, such that after each annihilation, the vortices are reintroduced at random  independent positions within the disk, ensuring that the total number of point vortices and $\eta$ remain constant.

\section{Equilibrium description of the imbalanced vortex gas}
\label{sec:equilibrium}

\noindent 
We recall Onsager's argument for negative temperature states, and the Joyce-Montgomery MF description of point-vortex equilibria. We then discuss transitions between zonal and dipolar flows in the so-called ``strong mixing limit'', obtained by linearizing the Joyce-Montgomery theory about the uniform state.  
 
\subsection{Onsager's microcanonical ensemble}
\label{sec:onsager}
\paragraph{Microcanonical density of states.}
The conservation of energy, angular momentum and imbalance under the Hamiltonian dynamics described in \S\ref{sec:hamiltonian} motivates a microcanonical description of the point-vortex statistical equilibria. A key feature  observed already by Kirchhoff~\cite{Kirchhoff1883Vorlesungen} is that the point-vortex coordinates themselves act as conjugate variables in Hamiltonian dynamics. Hence,  the accessible phase space  is  the Cartesian product of  the physical domain. Moreover, as the unit disk is bounded and  the number of vortices (and therefore dimensions) is finite, the phase space  has  a finite volume.
For the imbalanced point-vortex system, vortex configurations $\mathcal{C}$ are specified by $N$ vortex positions  $\br_1,\dots, \br_{N}$, with corresponding circulations $\Gamma_k = \pm 1$ for  $k=1,\dots,N,$ such that $N=N_++N_-$. The microcanonical density of states is then given by
\begin{equation}
	\Omega_N(E,M,\eta):=\dfrac{1}{N_+! N_-!}\int_{{\mathcal D}^N} \left(\prod_{i=1}^{N} d\br_i\right) \delta(\mathcal H [\mathcal C]-E)\,\delta(\mathcal M[\mathcal C]-M),
	\label{eq:micro}
\end{equation}
with the associated microcanonical entropy per vortex
\begin{equation}
	\label{fullS} s(e,\mu,\eta) := \dfrac{1}{N}\ln \Omega_N(N^2 e,N \mu,\eta).
\end{equation}
The factorial prefactors in~\eqref{eq:micro} account for  indistinguishability  of vortices within each circulation class. Here, $e=E/N^2$ and $\mu=M/N$ are the re-scaled energy and the angular momentum respectively. Such  thermodynamic scalings reflect the fact that the total energy scales linearly with the number of vortex pairs,  while the momentum scales linearly with the number of vortices~\cite{eyink1993negative}.\\

\paragraph{Negative temperatures.}
Let us briefly recall Onsager's argument for the existence of negative temperature states, which we here adapt to the microcanonical density  prescribed by Eq~\eqref{eq:micro}, incorporating constraints on  momentum and imbalance in addition  to energy.
 On the unit disk, the momentum  $\mathcal M_N[\mathcal{C}]$ is bounded, {ranging from $-N^-$ through $0$ (symmetric distribution) to $N^+$, with the two extrema values corresponding to a single-signed vortex population confined to the periphery of the disk and the opposite one clustered in the center. This implies that the rescaled angular momentum $\mu=\mathcal M_N[\mC]/N$ is bounded as
\beq
\dfrac{\eta-1}{2}\le \mu \le \dfrac{1+\eta}{2}.
\label{eq:etamu}
\eeq
On the other hand, under these conditions,   the point-vortex energy is unbounded, ranging from $-\infty$ to  $+\infty$. Large positive energies are obtained  when vortices of the same sign approach each other, whereas large negative energies arise when vortices of opposite signs approach each other or when a vortex approaches the boundary. At prescribed $\eta$ and $M=N\mu$, the cumulative phase-space volume corresponding to configurations  with energy less than $E$ 
\begin{equation}
	\Omega_N^{<}(E,M,\eta) = \int_{-\infty}^E  dE' \, \Omega_N(E',M,\eta), 
\end{equation}
increases monotonically from  $\Omega_N^<(-\infty,M,\eta)=0$ to $\Omega_N^<(+\infty,M,\eta) \propto \pi^{N}$.
This  implies that the density of states $\Omega_N(E,M,\eta) = \partial\Omega_N^</\partial E$ vanishes at both ends and  reaches at least one  maximum at some finite energy $E_m$.  By construction, this is also  an entropy maximum, characterized by  the infinite temperature 
\begin{equation}
	T= 
	\left(\dfrac{\partial s(e,\mu,\eta)}{\partial e}\right)^{-1}.
    \label{cumS}
\end{equation}
 Positive-temperature states correspond to low-energy equilibria $E<E_m$ where one expects the formation of vortex--antivortex dipoles.
By contrast, states with $E>E_m$ are  associated with  negative temperatures  and describe the clustering of same-sign vortices together. 
The  specific topology of the negative-temperature states is however not  prescribed by Onsager's argument, and is determined by
 the combined constraints imposed by  energy,  imbalance and angular momentum conservation.  Numerically, both the temperature and the typical vortex configurations can be accessed through microcanonical sampling; see Appendix~\ref{sec:MC}. At a theoretical level, this is the scope of the Joyce-Montgomery mean-field (MF) description.  
Note that  the MF theory is not relevant for the positive $T$ states dominated by vortex-antivortex dipoles,  because the spatial fluctuations surpass the mean-field effects. However, the remainder of the paper mostly focuses on the physics of negative-$T$ clusters. 
\subsection{Joyce-Montgomery mean-field theory}
\paragraph{Macrostates.} To describe the large-scale organization arising from the interplay between energy, imbalance and angular momentum, we rely on the MF approach introduced by Joyce and Montgomery~\cite{JoyceMontgomery1973}, and extended in~\cite{pointin1976statistical} to confined domains. Both works derive the  MF description by heuristically closing  the BBGKY hierarchy. We also highlight Ref.~\cite{chorin2013vorticity} for an elementary approach based on a counting argument and  Stirling's formula.
Joyce-Montgomery theory describes the microcanonical equilibria by coarse-graining the microscopic vortex positions into  smooth one-point  densities $n^+(\br)$ and $n^-(\br)$ which represent the  local densities of positive and negative vortices respectively,  with the macrostate $n(\br):= ( n^+,n^-)$ satisfying the normalization
\begin{equation}
	\bar \mC^\pm[n]= \int_\mD d \br\, n^\pm(\br) = \dfrac{N^{\pm}}{N},\quad\quad N^+ + N^-=N, 
	\label{eq:normalization}
\end{equation}
from which one constructs the coarse-grained vorticity and the associated streamfunction respectively as
\begin{equation}
	\bar \omega(\mathbf{r}) := n^{+}(\mathbf{r}) - n^{-}(\mathbf{r}),\quad \text{and }\quad\psi(\mathbf{r}) = \int_{\mathcal{D}}\,d\mathbf{r}'\,G(\mathbf{r}, \mathbf{r}') \, \bar \omega(\mathbf{r}'),
\label{eq:cg_fields}
\end{equation}
where $G(\br, \br_0) = -\Delta^{-1}\delta(\br-\br_0)$ is the Green's function in the unit disk $\mathcal{D}$ that vanishes at the boundary, as previously defined by Eq.~\eqref{eq:green}. 
Each macrostate  is characterized by its MF energy and angular momentum, respectively defined by
\begin{equation}
\bar \mH [n] = \frac{1}{2} \int_{\mD} d \br \, \psi(\br)\,
\bar \omega(\br),
\quad\text{and}\quad \bar \mM [n] = \int_\mD d \br \,\bar\omega(\br)\,|\br|^2.
\end{equation}
The imbalance parameter $\eta$ defined by Eq.~\eqref{eq:imbalance} enters through the total circulation
\begin{equation}
	\bar \mC[n] =  \int_{\mD} d \br \ \bar \omega(\br) = \dfrac{N^+-N^-}{N} = \eta,
\end{equation}
which is fixed once $N^\pm$ are prescribed.\\
\paragraph{Mean-field equilibria.}
MF equilibria are determined by identifying the most probable macrostate with prescribed macroscopic imbalance, energy,  
and angular momentum.
In the large-$N$ limit, this amounts to  maximizing the Gibbs entropy over admissible macrostates,
\begin{equation}
\begin{split}
&	S_{Gibbs}[n] := -\int_{\mathcal{D}} d\mathbf{r}\,\left(n^{+} \ln n^{+} + n^{-} \ln n^{-}\right) \to \sup,\\
& \text{subject to } \bar \mH [n] = e,\quad \bar \mM [n] =\mu,\quad   \quad \bar \mC^\pm [n] =\dfrac{1}{2}(1\pm \eta);
\end{split}
\label{eq:maxS}
\end{equation}
The macroscopic constraints involve the rescaled energy and momentum $e=E/N^2$ and $\mu = M/N$.
The resulting equilibrium macrostates are the  Gibbs densities 
\begin{equation}
	n_*^{\pm}(\mathbf{r}) = \exp\left(\mp x  - \gamma^{\pm}  \right),\quad x:= \beta  \psi_*(\mathbf{r})\,+  \alpha r^2 
\label{eq:gibbs}
\end{equation}
where the equilibrium streamfunction $\psi_*$ satisfies
the nonlinear  Poisson–Boltzmann  (PB)  equation:
\begin{equation}
\Delta \psi_* =-\bar \omega_*,\quad\bar  \omega_* (\br)= n^+_* -n_*^-.
\label{eq:PB}
\end{equation}
Together with the four constraints appearing in the variational problem~\eqref{eq:maxS}, Eqs.\eqref{eq:gibbs}-\eqref{eq:PB} form a closed system that determines the equilibrium macrostate. 
The resulting macrostate  entropy is 
\beq
	s_*(\beta,e,\mu)=S_{Gibbs}[n_*]:=  2\beta e+ \alpha  \mu + \dfrac{1}{2}(\gamma^++\gamma^-) + \dfrac{\eta}{2} (\gamma^+-\gamma^-)
	\label{eq:Sstar}
\eeq
The coefficients   $\alpha$, $\beta$, $\gamma^\pm$ are nonlinear functions of the constraints $e,\mu, \eta$, and determine the flow morphology. They are in general only known implicitly, and, depending on the invariants, need not be unique~\cite{pointin1976statistical}.
Physically,  $\bar \omega_*$ is a  stationary solution of the Euler equation in a frame rotating with angular velocity $\Omega=-2\alpha/\beta$. 

\subsection{Mean-field transitions between zonal and dipolar equilibria}
\label{sec:mfeq}
\paragraph{Fourier-Bessel modes.}
To characterize the topologies of the MF equilibria represented by the PB equation~\eqref{eq:PB}, we introduce the Fourier-Bessel decomposition
\begin{equation}
    \psi_*(\br)=\sum_{m\in\mathbb{Z}}\sum_{l\geq 1} p_{lm}\phi_{lm}(\br),\quad\text{where}\quad p_{lm}:=\scal{\psi_*,\phi_{lm}}.
	\label{eq:psi_fb}
\end{equation}
and hereafter use the shorthand
\beq
	\scal{f,g}= \int_\mD d \br\,f(\br)\,g^{*}(\br),
\eeq
with  the superscript * denoting complex conjugation. The modes $\phi_{lm}$  are explicitly given by
\begin{equation}
\phi_{lm}(\br)=\frac{e^{im\theta}}{\sqrt{\pi}}\frac{J_{|m|}(j_l^{(|m|)}r)}{|J_{|m|+1}(j_l^{(|m|)})|}, \quad\text{for}\quad \br = (r,\theta).
\end{equation}
Here,   $J_m$ is the Bessel function of the first kind of order $m$ (after $m\in\mathbb{Z}$), and $j_l^{(m)}$ denotes its $l$-th zero ($l\in\mathbb{N}$).
The modes $\phi_{lm}$ form an orthonormal basis of eigenfunctions for the Laplacian operator, \emph{e.g.} they satisfy $\scal{\phi_{lm},\phi_{l'm'}} = \delta_{ll'}\delta_{mm'}$, and diagonalize the Laplacian operator with eigenvalues $-\lambda_{lm}^2$ as
\begin{equation}
-\Delta \phi_{lm} := -\left(\frac{\partial^2 }{\partial r^2}
+\frac{1}{r}\frac{\partial }{\partial r}
-\frac{m^2}{r^2}\right)\phi_{lm}
= \lambda_{lm}^2\,\phi_{lm},\quad\text{with}\quad \lambda_{lm}:=j_l^{(|m|)}.
	\label{eq:eigen}
\end{equation}
Eqs.~\eqref{eq:PB}, \eqref{eq:psi_fb} and \eqref{eq:eigen} therefore imply the vorticity decomposition as
\begin{equation}
    \bar{\omega}_*(\mathbf{r})= \sum_{m\in\mathbb{Z}}  \sum_{l\geq 1} p_{lm}\lambda^2_{lm}\, \phi_{lm}(\br).
	\label{eq:w_fb}
\end{equation}
To ensure that $\bar{\omega_*}$ and $\psi_*$ are real-valued, the coefficients are Hermitian symmetric, \emph{e.g.}  $p_{lm} = p_{l{,-m}}^*$.
\paragraph{Strong mixing limit.}
Solving the maximization problem \eqref{eq:maxS}
 in full generality to track  the bifurcations  between various flow morphologies is  known to be cumbersome. For this reason, we here restrict our analysis to the so-called ``strong mixing limit'' discussed in \cite{Chavanis2014StatMech2DPointVortices}, which corresponds to  high-temperature (weakly non-uniform) states. This much simpler case is obtained by  linearizing the macrostate near the uniform background in  the rotating reference frame---in practice using   $x=\beta\psi + \alpha r^2 $ as a small parameter.  Note that both terms in $x$ must be small separately because $x$ should remain small for both 
 $r=0$ (when the second term vanishes) and $r=1$ (when the first term vanishes).
 
The linearisation of the macrostate $n^\pm$ yields
\beq
	n_*^{\pm}(\mathbf{r}) = \left(1\mp x_*\right) \exp\left(- \gamma^{\pm}  \right),\quad x_*= \beta  \psi_*(\mathbf{r}) +  \alpha r^2 
\eeq
along with the linearized PB
\beq
-\Delta\psi_* + B\psi_*=C  -Ar^2 ,\quad D:=e^{-\gamma^+}+e^{-\gamma^-},\quad  C:=e^{-\gamma^+}-e^{-\gamma^-} ,\quad B=D\beta,\quad A=D\alpha.
\label{eq:linH}
\eeq
Equation~\eqref{eq:linH} is an inhomogeneous Helmholtz equation. Like the original PB, it has to be solved with the boundary condition $\psi_* =0$ at $r=1$. As  usual for linear inhomogeneous ODEs, the general solution consists of a particular solution plus the general solution of the homogeneous part. The latter is the null space of the Helmholtz operator,
which is not empty only for discrete set of values of B corresponding to the zero eigenvalues, i.e. $B_{lm} = -\lambda^2_{lm}.$
Given this general representation, 
the main  difficulty is to solve Eq.~\eqref{eq:linH} in a microcanonical manner. This requires determining the coefficients $A,B,C,D$ compatible with  prescribed values for the energy, the angular momentum and the imbalance. We refer to Appendix \ref{sec:linear} for details. 
In brief, and away from the degenerate case $\mu =\eta /2$, the solutions belong either to axisymmetric 
branches 
or to  mixed branches occuring at non-axisymmetric modes. 
Axisymmetric branches are branches of particular radial solutions,  which exist at any $B$ from a continuous range and which we hereafter refer to as ``continuum'' branches.
Mixed branches are obtained by  superposing a non-axisymmetric mode to a continuum background, hence their names.
Specifically, at fixed $(e, \eta, \mu)$,  the stream function is solved as
\beq
\psi(e,\eta,\mu) = 
\left\lbrace
\begin{split}
	& \psi_c(B_*,\eta,\mu) \quad \text{where $B_*$ is such that }  e =\mathcal E_c(B_*,\eta,\mu) \quad \text{(continuum)},\\
	& \psi_c(-\lambda_{l_0m_0}^2,\eta,\mu) + 2 \Re( \chi \phi_{l_0m_0}) \quad\text{ where } |\chi|^2 =\dfrac{e-  \mathcal E_c(-\lambda_{l_0m_0}^2,\eta,\mu)}{\lambda^2_{l_0 m_0}} \quad m_0 \neq 0 \quad \text{(mixed)},\end{split}\right.
\label{eq:genpsi}
\eeq
where the functions $\psi_c(B,\eta,\mu)$ and $\mathcal E_c(B,\eta,\mu)$ have explicit expressions; see Appendix \ref{sec:linear}.
When several branches coexist at the same values of $(e,\mu,\eta)$, the optimization problem  \eqref{eq:maxS} selects the one which maximizes the macrostate entropy given by  Eq.~\eqref{eq:Sstar}; see Fig.~\ref{fig:2}.
\begin{figure}[H]
\centering
\includegraphics[width=0.94\textwidth]{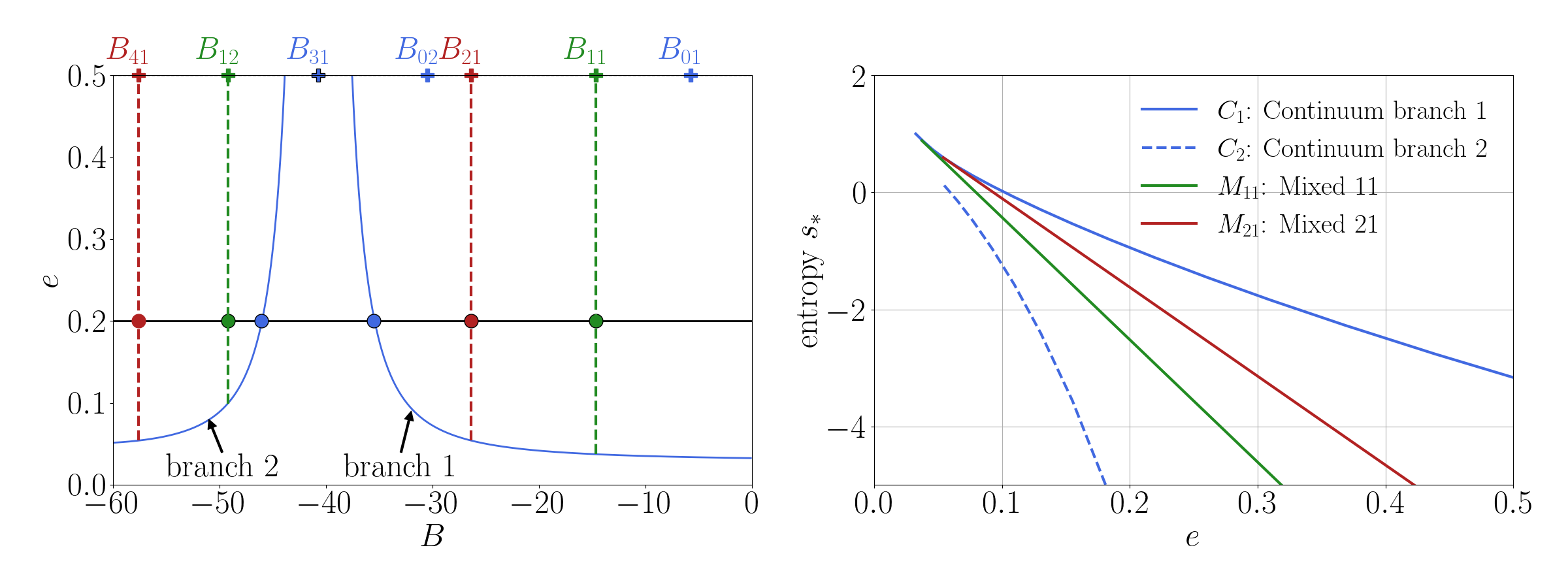}
\includegraphics[width=0.75\textwidth]{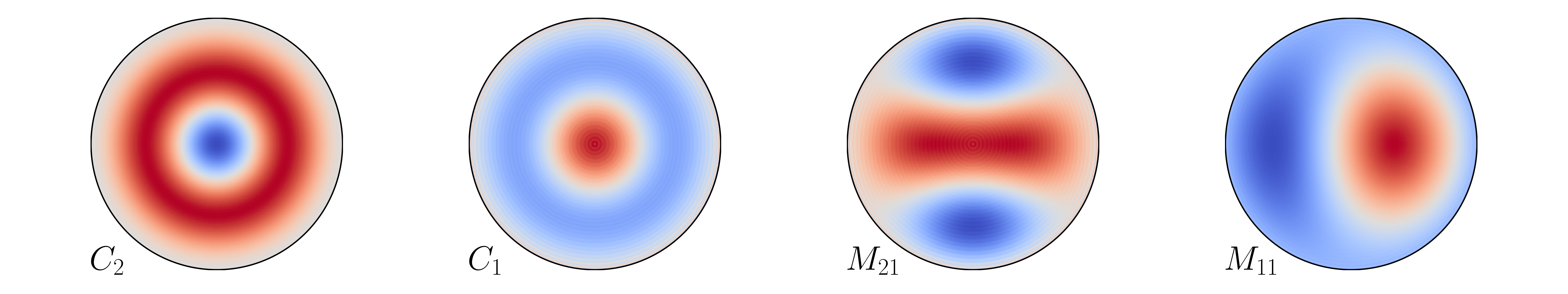}
\caption{Top left: Energy of the continuum branch  $\mathcal E_c(B,\eta,\mu)$, at fixed imbalance $\eta=0.2$ and angular momentum $\mu=-0.2$ as a function of $B$ (solid blue). The dashed vertical lines represent branches of mixed solutions, with the values on top  indicating $B_{ml} = -\lambda_{ml}^2$. 
The horizontal line indicates a prescribed value $e=0.2$; this selects  solutions on each  branch (dots).
Bottom: Streamfunctions in the  frame rotating at $\Omega=-2\alpha/\beta$ for the two continuous ($C_1$ and $C_2$) and two mixed solutions  ($M_{11}$ and $M_{12}$). Red and blue indicate anticlockwise and clockwise motions respectively.
Top right: Macrostate entropy as a function of the energy for the first 4 families, as indicated.}
\label{fig:2}
\end{figure}
\paragraph{Numerical construction of the bifurcation diagram.}
For each set of values  $(e,\mu,\eta)$, we compute the candidate solutions and determine the one with the largest entropy. As shown in Fig.~\ref{fig:2}, the entropy maximum can be realized on  
one of the infinite number of branches parameterized by different values of the parameter $B$. We here restrict our analysis to the first five families: namely three axisymmetric continuum branches representing zonal flows and the first symmetry-breaking mixed solutions corresponding to $(m,l) = (1,1)$ (dipole) and $(2,1)$ (quadrupole). 
Computing the entropy over a finite energy band, we detect the maximal-entropy branch.  The higher-order branches are guaranteed to have lower entropies for the part of the phase space considered in this paper. The resulting phase diagrams are summarized in Fig.~\ref{fig:3}. 
For all values of the imbalance parameter, the dipoles (green region) dominate when the energy and the momentum are close to $\alpha=\beta=0$ corresponding to well-mixed quasi-uniform state. Away from this regime,  the boundaries separating the dipolar and the zonal regions (in white)   exhibit a complex and nontrivial structure. As $\eta$ increases, the dipolar region evolves from a symmetric shape occupying a range $ |\mu|\lesssim 0.2$ to an  asymmetric shape extending  towards  higher momenta. 
We note also that non-vanishing imbalance produces a region (light blue) where the zonal flow is determined by the second branch of the continuum, corresponding to a flow reversal; see the flow patterns in Fig.~\ref{fig:2}.
Also note  that the quadrupole mode corresponds to a tiny region of the phase space which is vanishing for higher values of $\eta$, and the third continuous branch does not show up at all.
For $\mu<\eta/2$, the global rotation  frequency $\Omega=-2\alpha/\beta$   is negative at small $e$ and large $e$. 
 It changes sign  on the line $\alpha =0$ and is constant upon increasing $e$  on the mixed branch.  For $\mu>\eta/2$, the situation is reversed, with the global rotation frequency $\Omega=-2\alpha/\beta$  at large $e$ being positive. On the line $\beta=0$, the rotation is undefined. 
 In this case, though,  the states are axisymmetric and 
 are therefore stationary states of the Euler dynamics in any uniformly-rotating reference frame.
\begin{figure}[H]
\centering
\begin{minipage}{0.8\textwidth}
\includegraphics[width=0.5\textwidth]{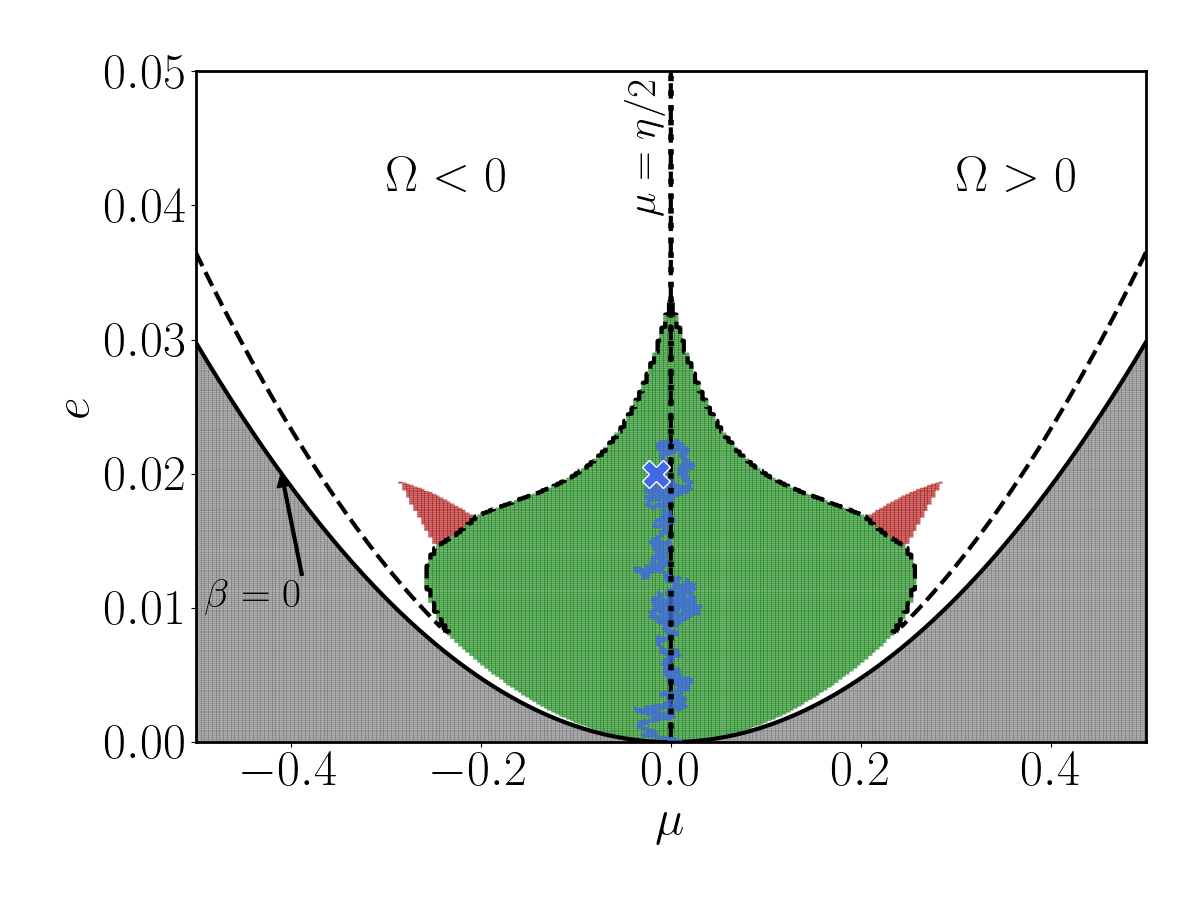}
\includegraphics[width=0.5\textwidth]{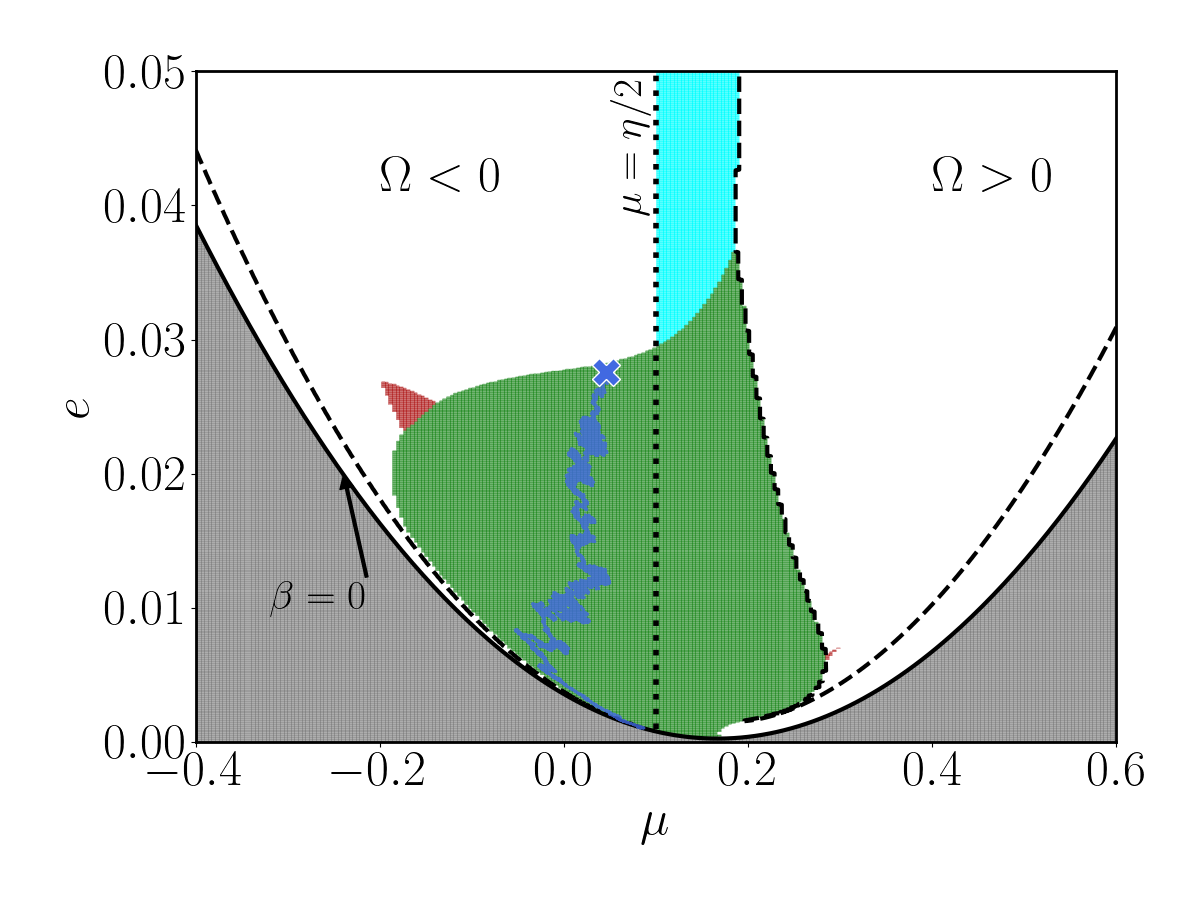}\\
\includegraphics[width=0.5\textwidth]{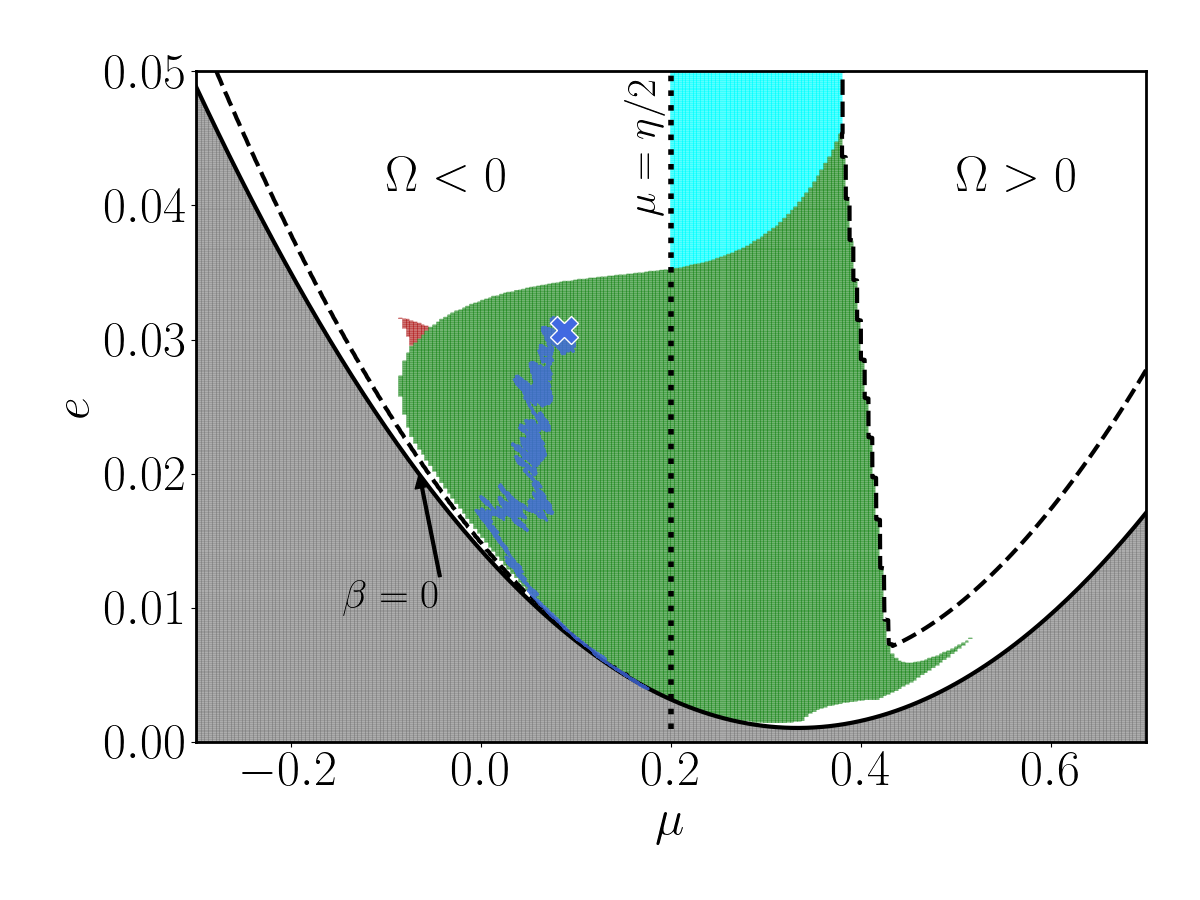}
\includegraphics[width=0.5\textwidth]{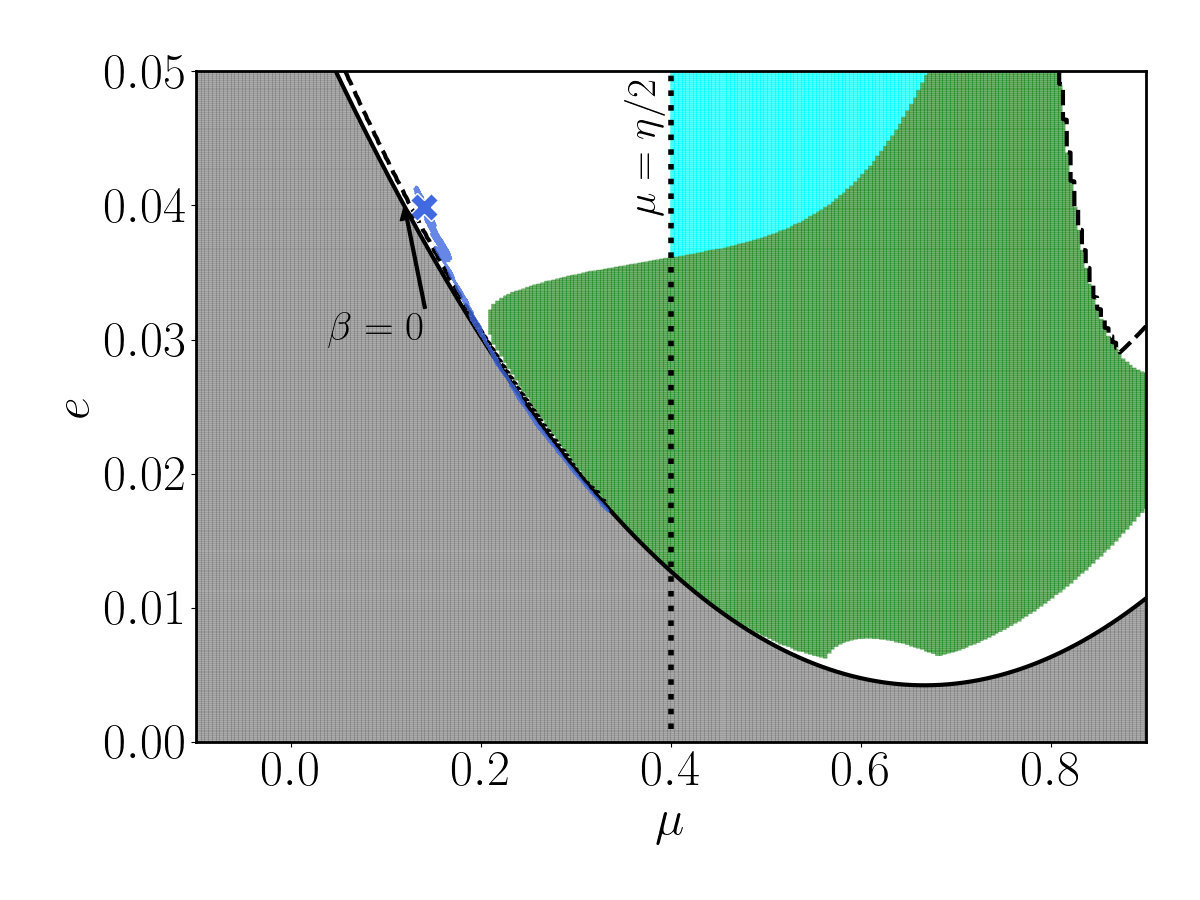}
\end{minipage}
\caption{Maximal-entropy branches computed for various $\eta=0,0.2$ (top) and $\eta= 0.4,0.8$ (bottom) from  left to right. The diagrams are computed over the admissible momentum range  $\mu \in (\eta/2-1/2,1/2+\eta/2)$ and the energy range $e \in (0,0.05)$. Colors indicate the entropy-maximizing branch: either branch 1 or 2 of the  continuum (C) corresponding to zonal flows, or   mixed solutions (M). The mixed solution at $(m,l)=(1,1)$ corresponds to dipolar flows (green) and the one at $(2,1)$ (red)  to  quadrupolar flows,   the white and cyan regions correspond to the continuum solutions $C_1$ and $C_2$ respectively.  The grey regions indicate the positive $\beta$ region. The black dashed  lines denote $\Omega=0$. 
The blue wiggly lines correspond to the data of \S\ref{sec:dns}. 
}
\label{fig:3}
\end{figure}

\section{Non-equilibrium dynamics of the imbalanced vortex gas}
\label{sec:dns}
\noindent
We  now  discuss the imbalanced vortex gas subject to the AR mechanism. We report a transition between zonal and dipolar flows, which we analyze  both from dynamical  and statistical perspectives.

\subsection{Qualitative overview}

\paragraph{Numerical methods.} To simulate the point-vortex gas evolving under the  AR mechanism of \S\ref{subsec:FDS}, we integrate the Hamiltonian dynamics~\eqref{eq:dynamical_equations} using a fourth-order Runge–Kutta scheme. At each time step, we  compute the minimum distance between  oppositely signed vortices, as well as the minimum distance  between a vortex and its image. Whenever either distance falls below  the annihilation length $\varepsilon$, the corresponding pair is removed and reinjected as described in \S\ref{subsec:FDS}. We typically set the annihilation length to  $\varepsilon = d/10$, with $d = \sqrt{\pi / N}$ being the mean inter-vortex spacing. Unless otherwise stated, we specifically discuss simulations employing $N = 1000$ point vortices and imbalance values ranging from $0$ to $1$.
Simulations are initialized from random statistically uniform vortex configurations corresponding to an infinite-temperature state with finite uniform rotation.\\

\paragraph{Large-scale topology.} After a sufficiently long time, the point-vortex gas settles into the configurations shown in Fig.~\ref{fig:DNS}. The corresponding energies and momenta at the final simulation times are reported in Table~\ref{table:EM}.
\renewcommand{\arraystretch}{1.3}
\begin{table}[htp!]
\centering
\begin{tabular}{|l|c|c|c|c|c|c|}
\hline
Imbalance $\eta$ & 0.00 & 0.20 & 0.40 & 0.50 & 0.60 & 0.80 \\
\hline
Energy $e=E/N^2$ & 0.020 & 0.028 & 0.031 & 0.034 & 0.035 & 0.040 \\
\hline
Angular momentum $\mu=M/N$ & -0.016 & 0.046 & 0.087 & 0.065 & 0.086 & 0.141 \\
\hline
\end{tabular}
\renewcommand{\arraystretch}{1}
\caption{Values of the energy $E$ and angular momentum $M$ at final time  $t=10^5$ of typical realizations for the point-vortex gas under AR with $N=1000$ and using different imbalance values $\eta$.}
\label{table:EM}
\end{table}
Despite persistent microscopic fluctuations, the states exhibit very stable large-scale  topologies, suggesting that the dynamics have reached a statistically stationary state. The resulting stationary  topology shown in Fig.~\ref{fig:DNS} depends on the imbalance parameter.
With no imbalance, the gas forms a symmetric large-scale dipole.
As the imbalance increases from $\eta=0.2$ to $0.5$, the dipole structure persists, but gradually loses its mirror symmetry. 
At $\eta=0.5$,  a small negative cluster is embedded within  a large-scale zonal flow of positive vorticity. It remains, however, strong enough to oppose the background  shear and maintain its self-rotation. This configuration visually resembles the ``Great Red Spot" vortex embedded in Jupiter's zonal flow~\cite{BouchetSommeria2002JFM}.
At $\eta=0.6$, an asymmetric  zonal flow  emerges, with no apparent region of closed streamlines formed by the negative vortices. At $\eta=0.8$, the  flow is nearly axisymmetric. 
\begin{figure}[htp!]
\centering
\includegraphics[width=\textwidth]{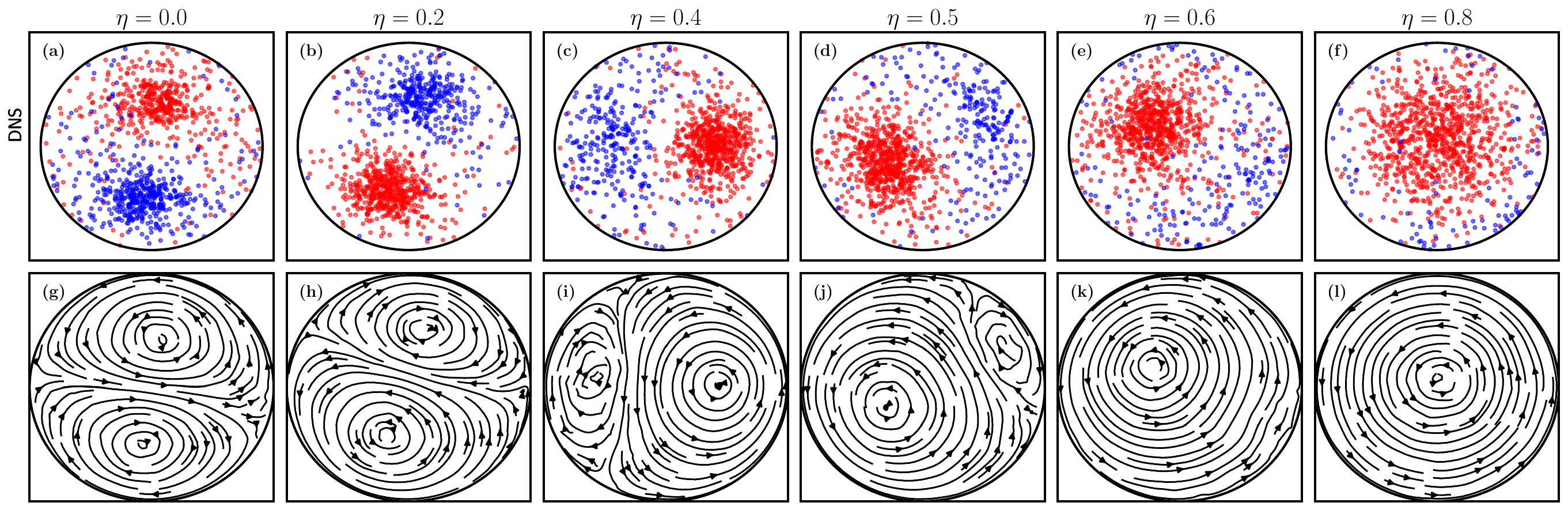} 
\caption{Final states obtained from the dynamical simulations of  Table \ref{table:EM}. The top row shows the vortex positions, whereas the bottom row plots the  streamlines of the corresponding flow.} 
\label{fig:DNS} 
\end{figure} 
\\
 
\paragraph{Energy and angular momentum.}
Despite the apparently stationary large-scale topology, the point-vortex gas under AR is not, strictly speaking, in  a statistically steady state since both the energy and the angular momentum  keep drifting slowly during the entire simulation window.
Figure~\ref{fig:energy} shows  the time evolution of the energy $E$ (left panel) and the angular momentum $M$ (right panel) for the simulations of Table~\ref{table:EM}. We see that, on average,
the energy increases with time for all the simulations. The reason is that upon mutual annihilation the system gains  net  energy which is not compensated by the reinjection mechanism, as the latter reinserts opposite-signed vortices at a distance which is of the order of the mean inter-vortex distance $d \gg \varepsilon$. As large-scale clusters  form, the  number of tight dipoles decreases, causing  the AR process to slow down.
\begin{figure}[htp]
\centering
\includegraphics[scale=0.48]{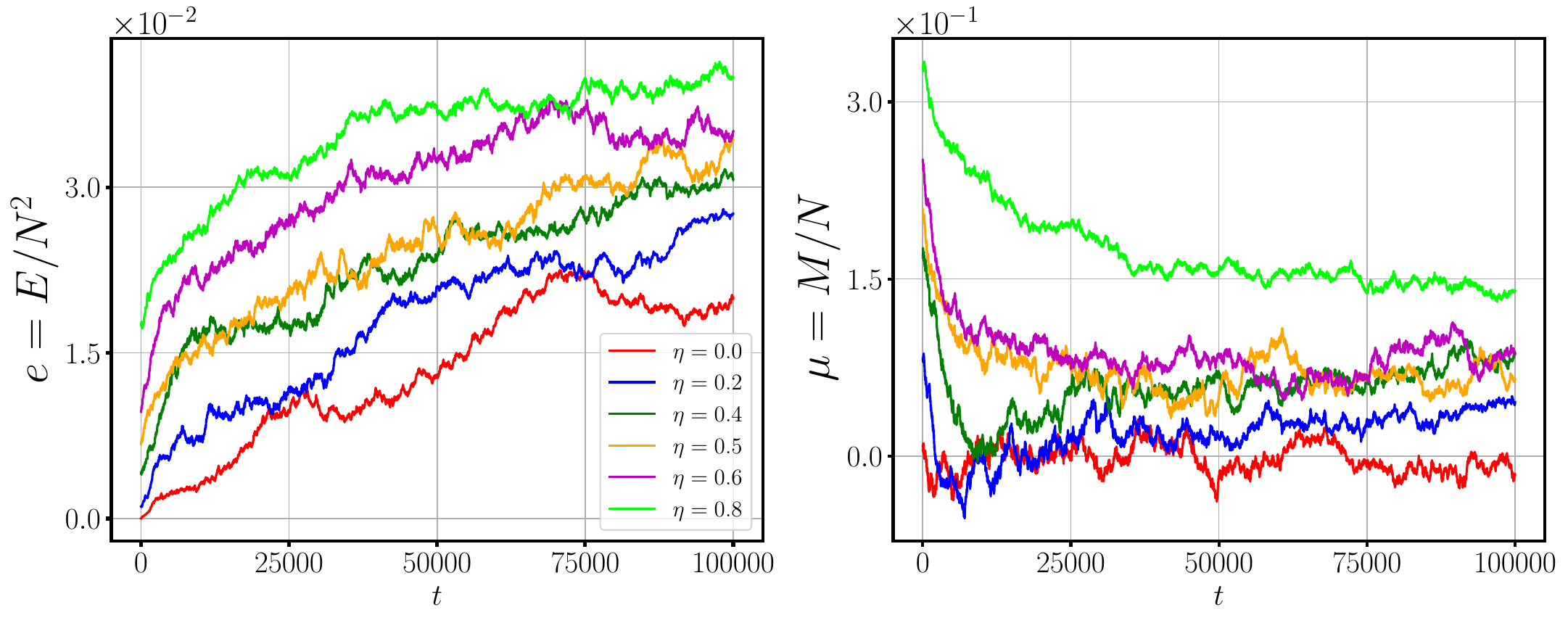} 
\caption{Time evolution of  the energy $E/N^2$ (left) and of the angular momentum $M/N$ (right) obtained from point-vortex simulations under the annihilation--reinjection mechanism for various imbalance parameters.
} \label{fig:energy} 
\end{figure}
While no systematic stabilization is observed, the energy  evolves at  a progressively slower rate as time increases, suggesting a  gradual saturation. This behavior is particularly apparent for the simulations at $\eta=0, 0.6$ and $0.8$, where the energy appears to  fluctuate around a nearly stationary average  for times $t>8000$. For large imbalance, the angular momentum shows a statistical trend for decrease. This indicates that the system moves from a uniform distribution to an energy-driven configuration where the dominant-sign vortices cluster near the center of the disk whereas the oppositely signed vortices are pushed away from the center.
For moderate imbalance ($\eta=0.2$ and $0.4$), we observe a short initial interval when $M$  drops abruptly before recovering.
This effect can be examined more closely in Fig.~\ref{fig:Meta04}, which correlates the time series of  the angular momentum  at $\eta=0.4$ with the corresponding vortex states.  Initially in Fig.~\ref{fig:Meta04}(a), the flow organizes into transient zonal structures. At later times in Fig.~\ref{fig:Meta04}(b), the zonal flow structure persists but becomes displaced from the center, while the angular momentum begins to increase. Eventually in Fig.~\ref{fig:Meta04}(c,d), as the angular momentum stabilizes and slowly increases,  the system settles into a stationary  dipolar configuration.
 \begin{figure}[htp!]
\centering
    \includegraphics[scale=0.55]{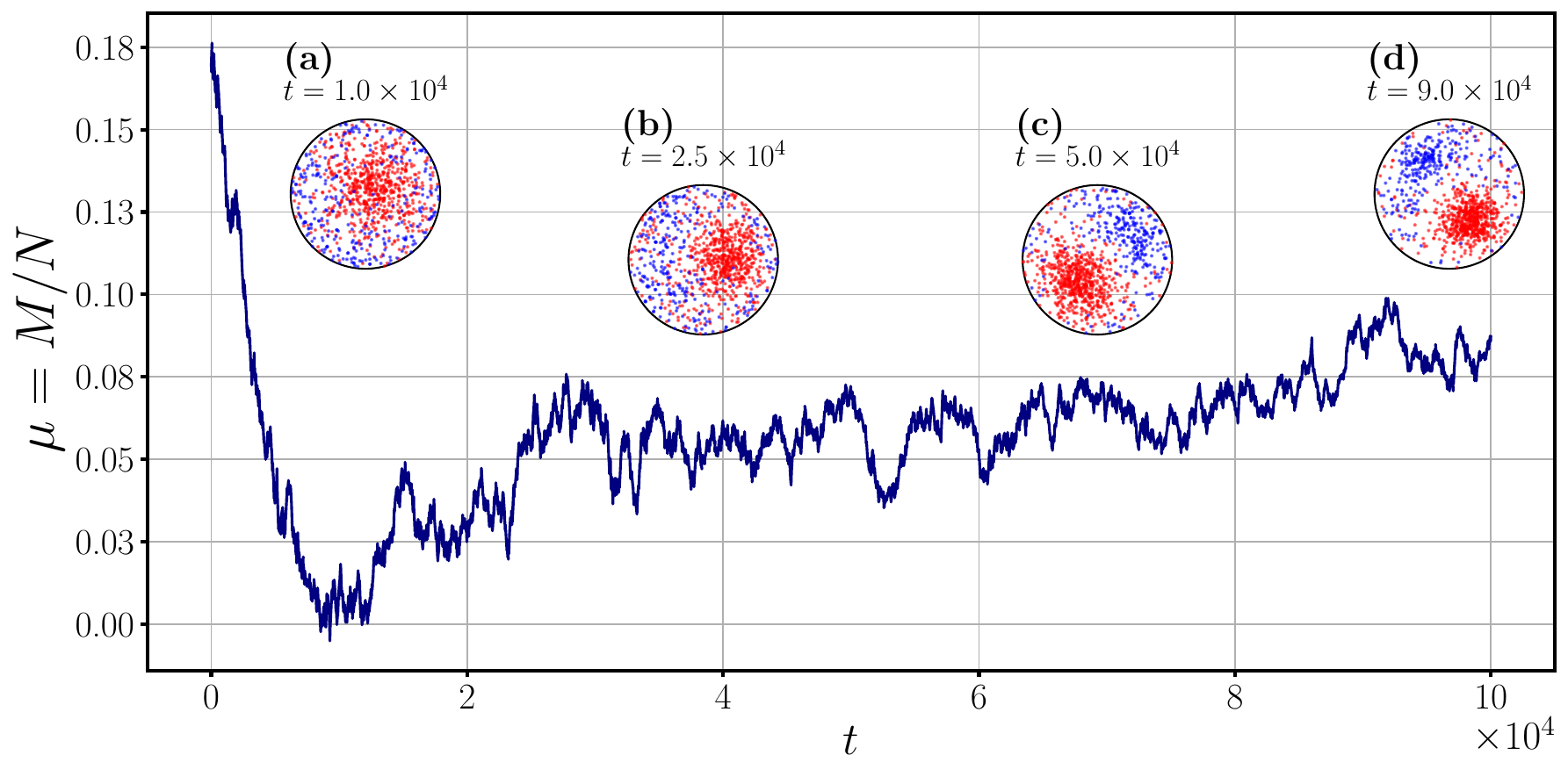}
    \caption{Time series of the angular momentum for the run reported in Table~\ref{table:EM}, corresponding to an imbalance of $\eta = 0.4$, together with selected instantaneous configurations at representative times.}
    \label{fig:Meta04}
\end{figure}

\subsection{Adiabatic evolution driven by annihilation--reinjection }
\noindent 
The transitions observed between zonal and dipolar flows can be interpreted in the light of the equilibrium phase diagram. This ultimately suggests that the forcing induces an approximately adiabatic evolution through the microcanonical equilibrium phase space.\\

\paragraph{Phase transition of the final state.} By construction, as the number of AR events decreases, we  could expect the dynamics to converge towards large-scale equilibrium.  We therefore compare the final states of the direct numerical simulations (DNS) of Table~\ref{table:EM}  with equilibrium predictions. 
In addition to  the MF solutions, we sample Onsager's microcanonical ensemble using the Creutz algorithm  described in Appendix~\ref{sec:MC} using as target energies and momenta the values reported in Table~\ref{table:EM}.
The resulting vortex configurations are shown in Fig.~\ref{fig:MCvsDNS}, showing a transition from dipolar to zonal flows as the imbalance  increases beyond $\eta \sim 0.6$ both for DNS and equilibrium predictions. This is consistent with Fig.~\ref{fig:3}, which shows the trajectories in the phase space $(e,\mu)$ for various values of $\eta$, together with the final state obtained from the DNS (represented by a cross). We observe that at low imbalance ($\eta\lesssim 0.6$), the final states lie inside the green region which corresponds to dipolar flows. To quantitatively compare the DNS with the  equilibrium theory, we introduce the dimensionless order parameters:
\begin{equation}
    \Lambda_{11}=\frac{|p_{11}|^2}{|p_{11}|^2+|p_{10}|^2}\quad\textup{and}\quad \Lambda_{10}=\frac{|p_{10}|^2}{|p_{11}|^2+|p_{10}|^2}.
    \label{lambda}
\end{equation}
Values of the parameter $\Lambda_{11}\simeq 1$ indicate the predominance of dipoles, while $\Lambda_{10}\simeq 1$ indicate the predominance of a zonal flow. Results are shown in Fig. \ref{fig:8}, again comparing the final states of the DNS with equilibrium predictions. We observe a good agreement between the DNS, MC, and linear MF theory. In particular, the linear theory provides a very accurate description of the topology of the final states; this is surprising given that the latter are strongly non-uniform and \textit{a priori} beyond the scope of the linear description.\\
   
\begin{figure}[htp!]
    \centering
    \includegraphics[scale=0.46]{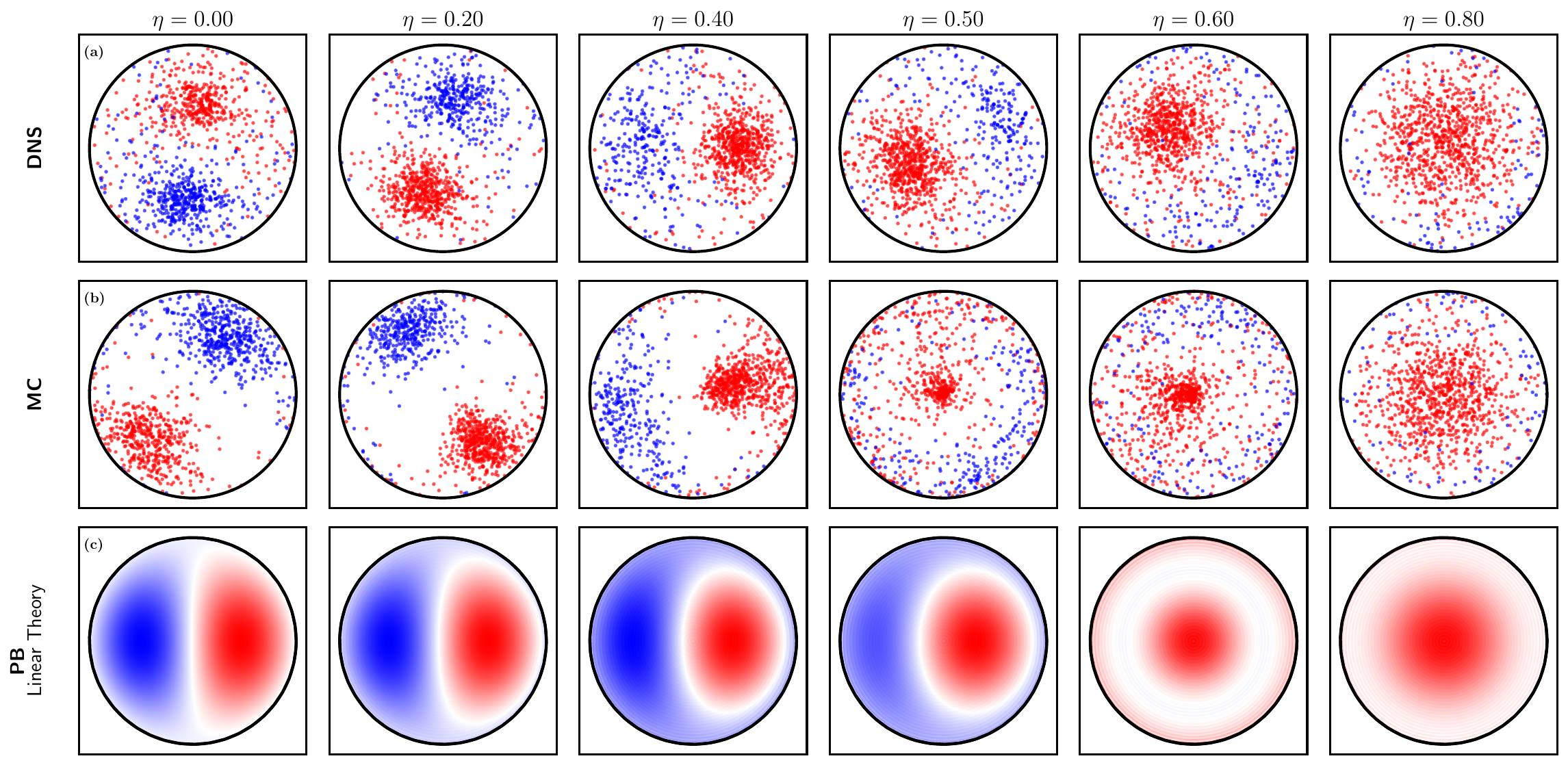}
    \caption{Vortex configurations observed from (a) AR DNS of Table~\ref{table:EM}, 
    (b) microcanonical MC simulations using  the values of $E$ and $M$ from Table~\ref{table:EM},
     (c) linearized MF solutions of Poisson-Boltzmann equation~\eqref{eq:PB} also using  the values of $e$, $\mu$ and $\eta$ from Table~\ref{table:EM}.}
    \label{fig:MCvsDNS}
\end{figure}

\paragraph{Bifurcation scenario.} The agreement between  the  linear theory and the DNS  can be understood by inspecting the temporal dynamics of the DNS rather than the final states alone.
Let us consider again  the case $\eta=0.4$, which corresponds to the time series shown in Fig.~\ref{fig:Meta04}.
Two dynamical regimes clearly appear. 
The first   is an initial relaxation ($t\lesssim 10^4$) where both the angular momentum and the energy  sharply vary, respectively decreasing  from $\mu \simeq 0.2$ to $\mu \simeq 0$ and increasing from $e=0.004$ to $e\simeq 0.016$.  Throughout this stage, the vortex configuration  remains quasi-uniform.
The second stage  is observed for $t\gtrsim 2\times10^4$: the energy continues  to increase, while the angular momentum  stabilizes around $\mu \simeq 0.05$.
The  flow is  dipolar, and this organization becomes increasingly pronounced as the energy rises. 

This dynamical behavior can be interpreted in the light of the linear equilibrium diagrams. The left panel of Fig.~\ref{fig:lambda} represents the dynamical trajectories in the $(e,\mu)$, superimposed on the equilibrium regions computed in Fig.~\ref{fig:3}. 
With this representation, we interpret the initial relaxation as a stage  where the system follows the $\Omega=0$ branch, constantly adjusting its angular momentum toward the most probable one as the energy  is increased. In this stage, the system is quasi-uniform, non-rotating, and therefore plainly in the scope of linear theory.
At $\mu=0$, the system exhibits a bifurcation, signaled by  the trajectories entering the dipolar region. This implies that the linear theory determines the bifurcation scenario between dipolar and zonal flows.

This dynamical scenario  is seen for the other values of imbalance.
In the linear equilibrium diagram of Fig.~\ref{fig:3}, the blue wiggly lines display the dynamical trajectories of the DNS in the corresponding $(\mu,e)$ plane. For $\eta \lesssim 0.6$, the phenomenology is similar to that described for $\eta=0.4$: the system first relaxes toward $\mu\simeq 0$ along the $\alpha=0$ line, and then bifurcates  toward the dipolar region.
For $\eta=0.8$, the relaxation stage persists, but the dipolar region   is no longer an equilibrium state in the vicinity of  $\mu=0$. The system therefore remains in a purely zonal, axisymmetric state. In all cases, the linear theory selects the relaxation trajectories towards $\mu=0$, and determines the subsequent emerging branch.
\\
\paragraph{Three observations.}
We conclude with three observations.
First, we note that  as the energy is increased, the system progressively drifts away from quasi-uniformity. Although the linear theory still qualitatively captures  the topology, we observe quantitative deviations; see the  right panel of Fig.~\ref{fig:8}. This is consistent with  non-uniform state becoming increasingly pronounced as the energy is increased.

Second, the dynamical behavior seems largely independent of both the annihilation distance and the number of vortices. Fig.~\ref{fig:lambda}             superimposes the trajectory of the reference case $N=1000,   \varepsilon=d/10$, with those obtained by varying  $\varepsilon$ and $N$. To characterize the flow configurations, we compute the values of $\Lambda_{10}$ and $\Lambda_{11}$ along the trajectories and  represent then  using color scales in the central and right panels. No dependence upon the specific trajectories is seen. Instead, the values of both parameters directly correlate to  the instantaneous values  of $e$ and $\mu$. This suggests that the system evolves quasi-adiabatically  across the microcanonical equilibrium phase space.

Third, observe that at high energies, some of the final states wind up outside of the dipolar region determined by the  linear equilibrium diagram. On the other hand,  no topological changes are reported for those trajectories, \emph{e.g.} the final states are still dipolar. This is compatible with the idea that  the linear theory only selects the bifurcation scenario: it applies when the system is  quasi-uniform and there it selects the branch that the system subsequently follows.
The subsequent dynamics reaches energy values for which the linear equilibrium is no longer  directly  relevant.

Potentially, there is also a possibility that the observed final states follow the dipolar branch whose entropy is no longer globally maximal but which is metastable. It  might then take a large fluctuation (and therefore time) to knock the system out of the local entropy maximum toward the global one realized on the zonal mode. However, we could not witness such  fluctuation  events in our simulations.

\begin{figure}[H]
    \centering
    \includegraphics[width=0.49\textwidth]{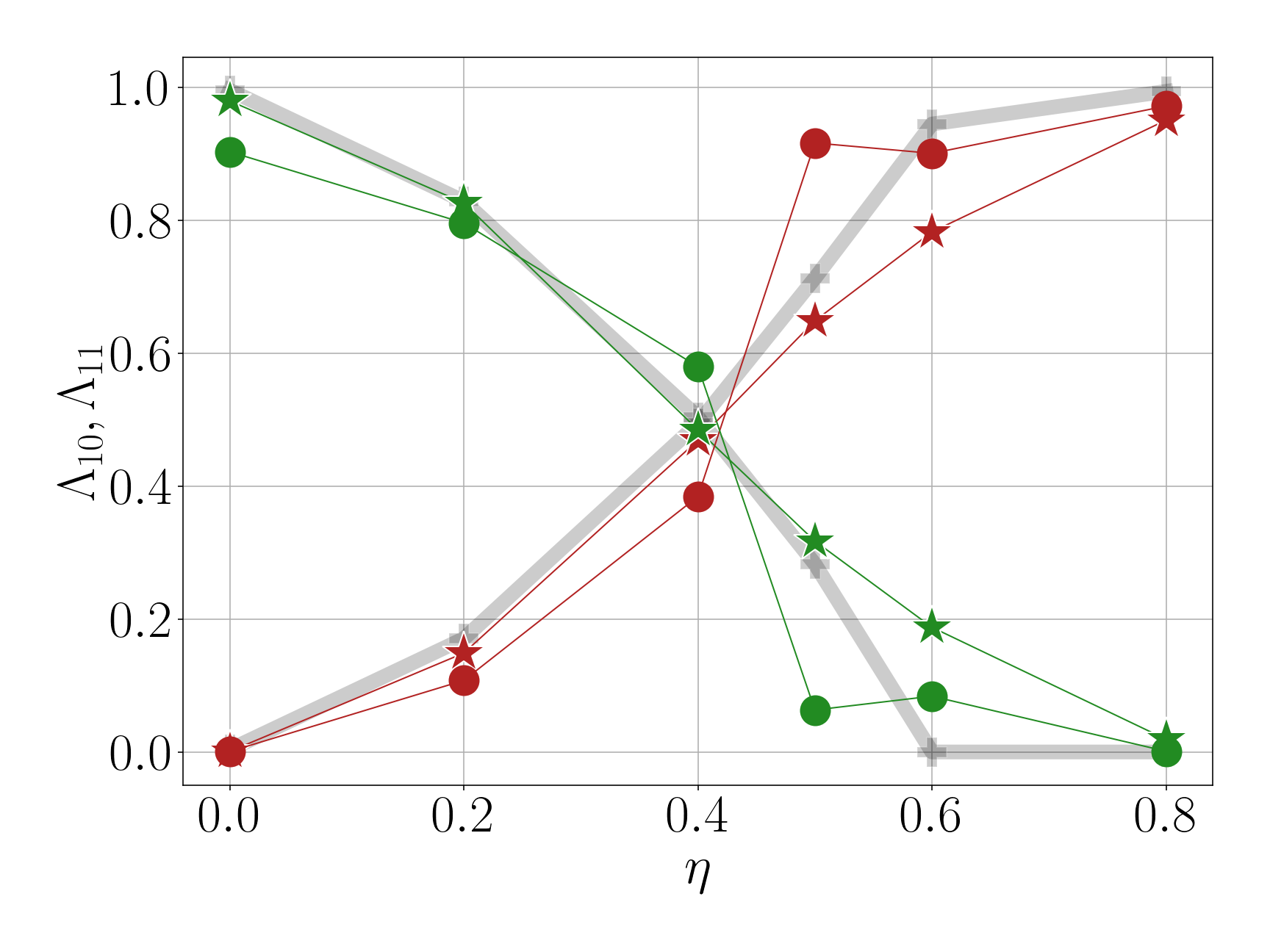}
\includegraphics[width=0.49\textwidth]{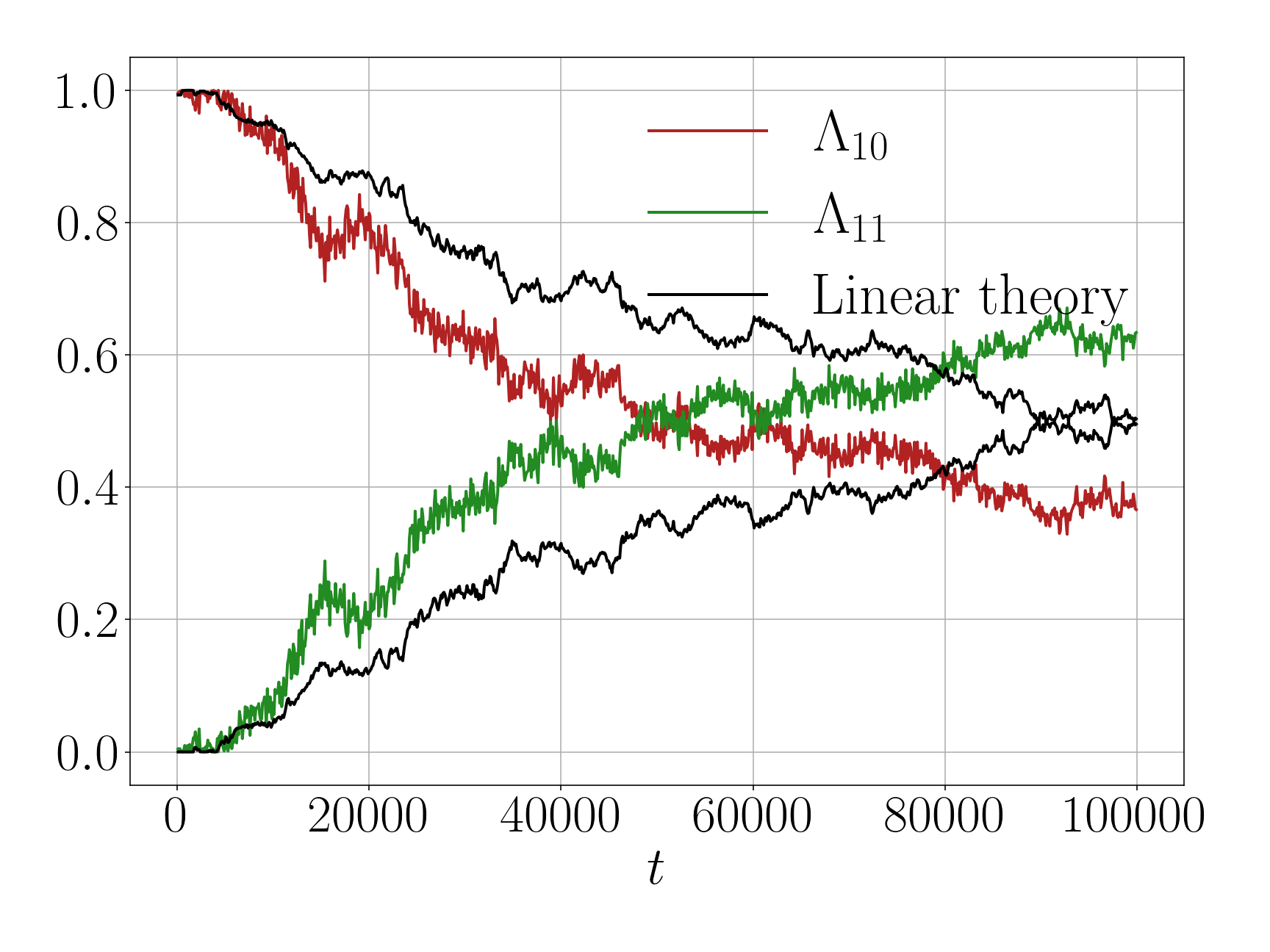}
    \caption{(Left) Parameters  $\Lambda_{10}$ (red) and $\Lambda_{11}$ (green) as a function of the imbalance for DNS at their final times ($\bullet$),  and respective MC simulations ($\star$) and linearized MF solutions (thick grey). Right: Time evolution of the parameters for $\eta=0.4$, compared with the  predictions from linear theory considering simulations for a system with parameters $N=1000$ and $\varepsilon/d=0.1$.
}
    \label{fig:8}
\end{figure}

\begin{figure}[H]
    \centering
    \includegraphics[scale=0.365]{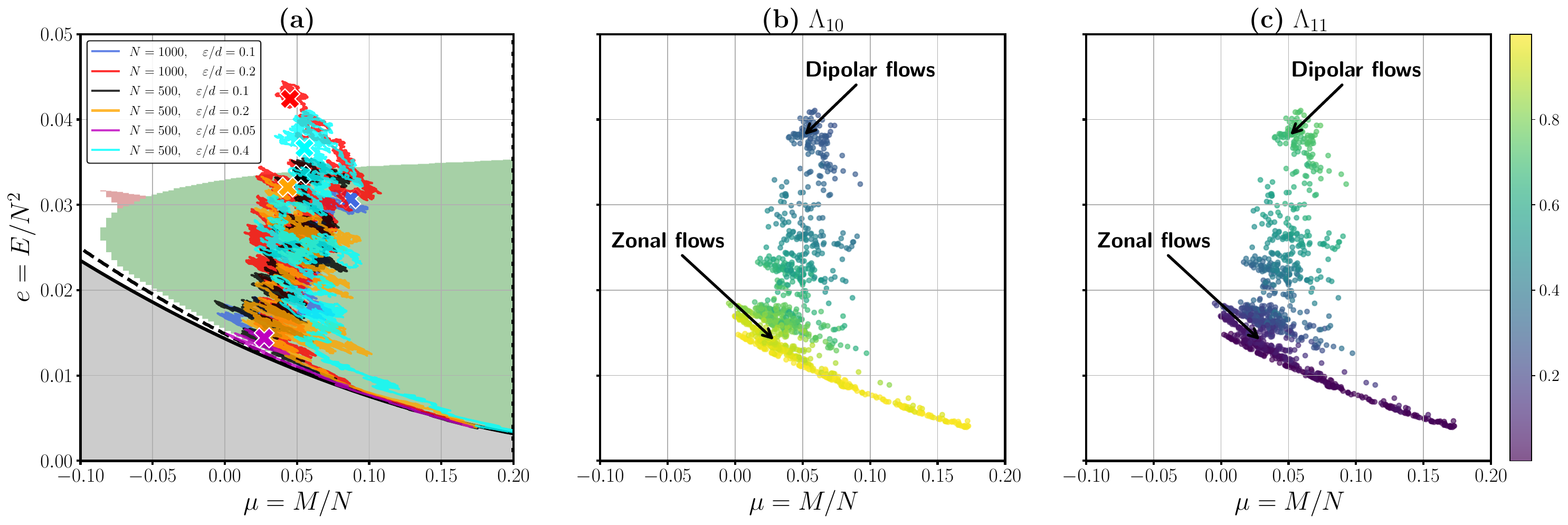}
    \caption{(a) Phase diagram in the $(e,\mu)$ plane, showing the corresponding DNS trajectories for different values of $N$ and $\varepsilon$ at fixed imbalance $\eta=0.4$. (b,c) Parameters $\Lambda_{10}$ and $\Lambda_{11}$ computed along the different DNS runs, indicating the regions of zonal and dipolar flow states. }
    \label{fig:lambda}
\end{figure}
\subsection{Shear-induced transition between zonal and dipolar states}
While the non-equilibrium AR dynamics generates a quasi-adiabatic evolution across the microcanonical phase space, the reason why the system bifurcates around $\mu=0.05$ cannot be readily  explained within the framework of statistical mechanics. We here describe a heuristic argument from classical hydrodynamics involving a shear instability.\\

\paragraph{A heuristic criterion for dipole destabilization.}
At a heuristic level, a dipolar flow is expected to break down when the shear generated by the dominant vortex cluster exceeds the self-rotation (vorticity) of the minority cluster. Consider a configuration where the positive vortices dominate and cluster near the center, while the negative  vortices are  clustered in an off-center spot; see Fig.~\ref{fig:transition}.  The flow shear  $s$ produced by the positive cluster at the position of the negative cluster can be estimated as $|s| \sim  N^{+}/R^2$. On the other hand, the mean vorticity in the negative cluster is $|\omega^{-}| \sim N^{-}/(d^{-})^{2}$, where $d^{-}$ is here the size of the large-scale negative vortex. The dipolar state becomes unstable when the external shear supersedes the  self-rotation  of the negative cluster, i.e., $|s| > |\omega^{-}|$, leading to
    $N^{+}/N^{-}>\left(R/d^{-}\right)^2$. In terms of the imbalance parameter and using $R=1$, the criterion for dipolar destabilization  becomes
\begin{equation}
    \frac{1+\eta}{1-\eta}>\left(\frac{1}{d^{-}}\right)^2.
    \label{eq:criterion}
\end{equation}
This heuristic argument supports the idea that a large imbalance in the number of positive and negative vortices favors zonal flows over dipolar clusters.  The transition also depends  on $d^{-}$, whose  value is not predicted by our hydrodynamic argument. 
\begin{figure}[htp]
   \centering
   \includegraphics[scale=0.98]{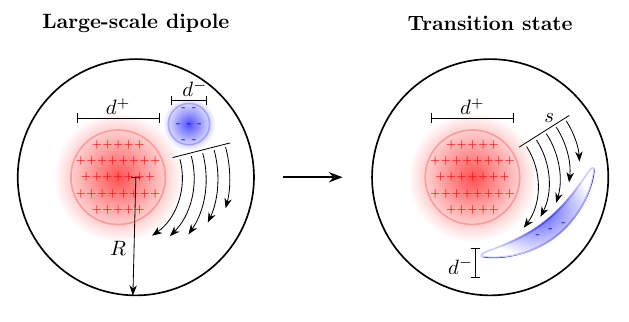}
   \caption{Destabilisation of a dipolar flow: The shear flow generated by the large-scale positive vortex stretches the smaller negative vortex.}
   \label{fig:transition}
\end{figure}
\\
\paragraph{Numerical tests.}
To test  the validity of the criterion given by Eq.~\eqref{eq:criterion}, we  estimate $d^{-}$ directly from the numerical simulations, and define it as the average pairwise distance between negative vortices:
\begin{equation}
d^{-}=\frac{1}{(N^{-}-1)N^-}{\sum^{N^{-}}_{\substack{j, k =1\\j\neq k}}}|\mathbf{r}_j^{-}-\mathbf{r}_k^{-}|.
\end{equation}
This determines the characteristic size of the set of negative vortices.
From the temporal evolution of the distance $d^{-}$, we identify when the criterion predicts that a zonal flow should form. Figure~\ref{fig:plot_criteria} shows the temporal dynamics of the ratio $\left(R/d^{-}\right)^2$, constituting the  right-hand side of Eq.~\eqref{eq:criterion}, and compares it with the left-hand side prediction, determined solely by the imbalance.
For zero imbalance, the criterion~\eqref{eq:criterion} is never satisfied, and as a matter of fact no zonal flow is observed. As the imbalance increases,  dipolar flows do not nucleate, and the criterion~\eqref{eq:criterion} is satisfied over increasing periods of time.
Note that the latter criterion  suggests a transition occurring at $\eta\simeq 0.4$. While the equilibrium theory rather identifies $\eta=0.6$ as the bifurcation value,  this is still consistent with  Fig.~\ref{fig:8} where 
 $\eta=0.4$ is the value at which dipolar and zonal modes contribute equally.
\begin{figure}[htp!]
    \centering
    \includegraphics[scale=0.46]{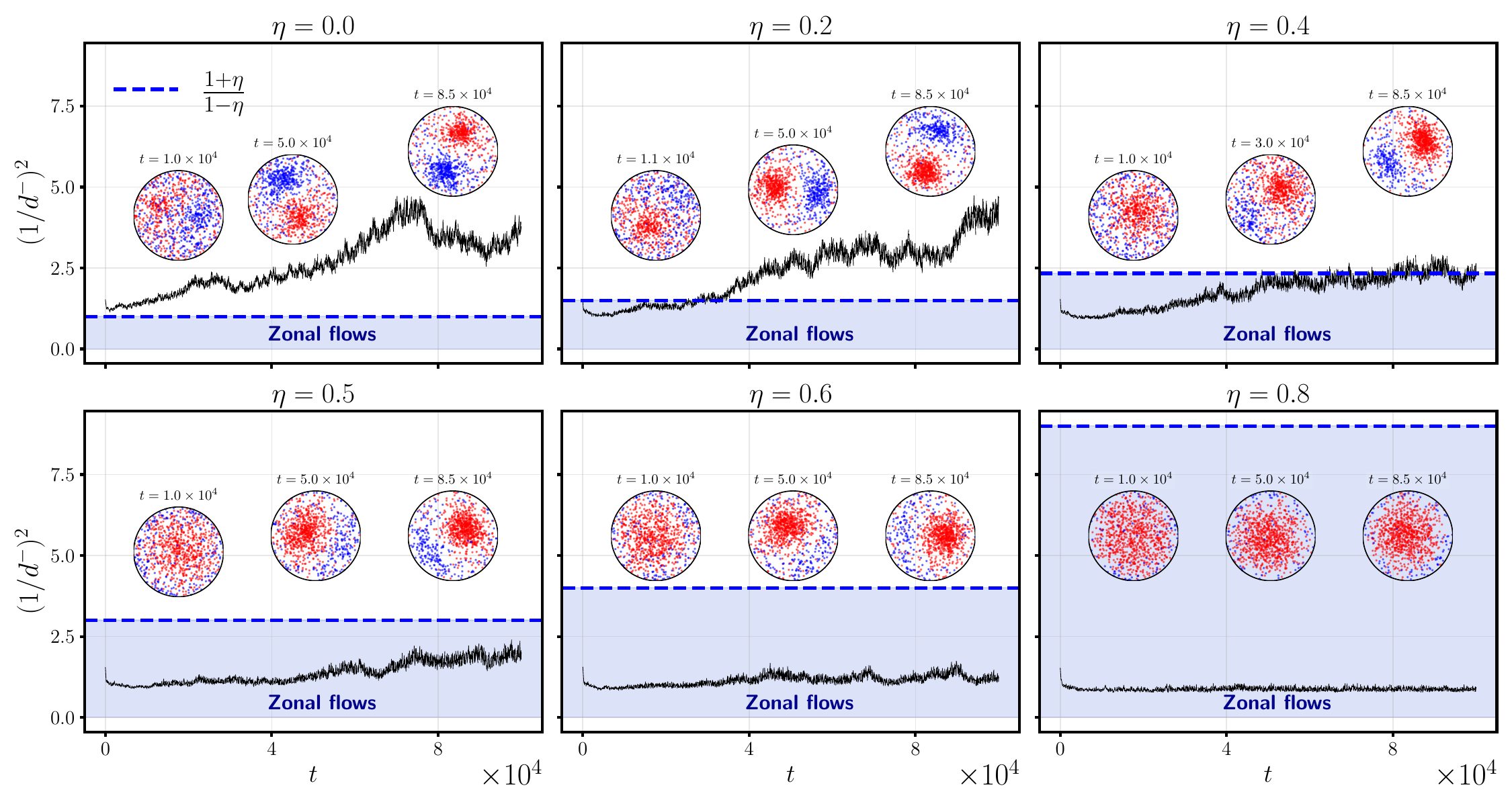}
    \caption{Temporal evolution of $(1/d^-)^2$  for different imbalance values and the runs of Table \ref{table:EM}. 
    The horizontal dashed green lines mark the threshold $(1+\eta)/(1-\eta)$. Whenever the threshold exceeds the black curve, the criterion in Eq.~\eqref{eq:criterion}  predicts destabilization of the dipolar state into  zonal flows.}
    \label{fig:plot_criteria}
\end{figure}

\section{Concluding remarks}
\label{sec:conclusion}
In this work, we have examined the emergence of zonal flows in a two-signed point-vortex gas evolving in a disk, emphasizing the role of the circulation imbalance $\eta$ in driving topological changes in the large-scale vortex configurations. We introduced a non-equilibrium extension of the point-vortex setup by supplementing the Hamiltonian point-vortex dynamics with an AR mechanism. This AR process acts as a crude representation of evaporative heating and optical forcing employed in BEC experiments.
In practice, it produces fluctuations in both the energy and the angular momentum, resulting in a statistical drift towards high-energy states. Over sufficiently long times the system reaches a statistically  stationary large-scale topology, reminiscent of Onsager's negative-temperature regime. The central observation is that the circulation imbalance drives a transition in the large-scale vortex topology.

For small imbalance, the flow is essentially a large-scale dipole. As the imbalance increases, the system evolves towards states of higher energy and angular momentum, and the dipolar configuration is eventually replaced by a  zonal-flow-dominated topology for $\eta \gtrsim 0.4$ and by a  pure zonal flow for $\eta \gtrsim 0.6$.
{From the hydrodynamic perspective}, this transition can be understood as a consequence of the shear flow generated by the dominant (say, positive) vortices becoming stronger than the vorticity associated with the negative vortices: this destabilizes the dipolar structure.
{From a statistical perspective}, the persistence of microscopic fluctuations at all times precludes, \emph{a priori},  the use of microcanonical equilibrium theory. 
It is, however, found that the transition in the dominant large-scale topology is reproduced by the direct MC sampling of Onsager's microcanonical measure, with prescribed values of the imbalance, energy, and angular momentum. 

Although the dynamics under AR is not microcanonical  nor  formally  at equilibrium, the close agreement between the DNS and the microcanonical MC results suggests that  the forcing induces a slow, approximately adiabatic evolution through the microcanonical equilibrium phase space.
From this perspective, the non-equilibrium nature of the AR setup is  primarily reflected in the specific trajectory followed through the ($e,\mu,\eta$) phase space. This trajectory cannot be predicted from equilibrium arguments, whereas the instantaneous large-scale states remain accurately described by equilibrium statistical mechanics.  
We provide a qualitative understanding of the microcanonical equilibria by examining the MF solutions of the Joyce--Montgomery theory. The numerical analysis of the corresponding Poisson--Boltzmann equation yields a phase diagram that reveals a nontrivial interplay between energy, angular momentum, and imbalance in the emergence of coherent flow structures. 
On the other hand, the agreement between topologies obtained from  microcanonical sampling   and AR DNS  suggests  that the effective heating with random reinjection acts as a  weak stochastic forcing, which restores ergodicity by promoting transitions between otherwise dynamically isolated regions of phase space. This mechanism arises  in the context of Langevin dynamics and theories of stochastic differential equations \cite{mattingly2002ergodicity}.

\color{black}
Beyond the questions specific to the point-vortex system, including the universality of the forcing, the statistics of the non-equilibrium trajectory, and finite-$N$ effects, we believe that our results  will be particularly relevant to superfluid two-dimensional turbulence consisting of point-like quantized vortices and waves, such as BEC turbulence realized in quasi-2D and bubble traps~\cite{GuoGutierrezReyBadrPerrin2022ExpansionShellTrap,reeves2013onsager}. When such traps are set into rotation, we conjecture that the vortex turbulence will self-organize into large-scale zonal flows once suitable values of the imbalance, energy, and angular momentum are realized. More generally, our results show that circulation imbalance, energy, and angular momentum must be considered jointly to understand coherent flow structures in non-equilibrium vortex turbulence. 

\section*{Acknowledgments}
 This work was supported by the French government through the France 2030 investment plan managed by the  French National Research Agency (ANR), as part of the Initiative of Excellence Université Côte d’Azur under reference number ANR-15-IDEX-01 and by the ANR project VORTECS (Grant No. ANR-22-CE30-0011). The authors are grateful to the Université Côte d’Azur’s Center for High-Performance Computing (OPAL infrastructure) for providing resources and support.

\bibliographystyle{apsrev4-2}
\bibliography{biblio.bib}

@article{BouchetSommeria2002JFM,
  author  = {Bouchet, F. and Sommeria, J.},
  title   = {Emergence of intense jets and {Jupiter}'s Great Red Spot as maximum-entropy structures},
  journal = {Journal of Fluid Mechanics},
  volume  = {464},
  pages   = {165--207},
  year    = {2002},
  doi     = {10.1017/S0022112002008789}
}

@article{bouchet2012statistical,
  title={Statistical mechanics of two-dimensional and geophysical flows},
  author={Bouchet, Freddy and Venaille, Antoine},
  journal={Physics reports},
  volume={515},
  number={5},
  pages={227--295},
  year={2012},
  publisher={Elsevier}
}

@article{creutz1983microcanonical,
  author  = {Creutz, M.},
  title   = {Microcanonical {Monte-Carlo} Simulation},
  journal = {Physical Review Letters},
  volume  = {50},
  pages   = {1411--1413},
  year    = {1983},
  doi     = {10.1103/PhysRevLett.50.1411}
}

@article{Chavanis2014StatMech2DPointVortices,
  author  = {Chavanis, P. H.},
  title   = {Statistical mechanics of two-dimensional point vortices: relaxation equations and strong mixing limit},
  journal = {The European Physical Journal B},
  volume  = {87},
  pages   = {81},
  year    = {2014},
  doi     = {10.1140/epjb/e2014-40869-x}
}

@book{chorin2013vorticity,
  title={Vorticity and turbulence},
  author={Chorin, A. J.},
  volume={103},
  year={2013},
  publisher={Springer Science \& Business Media}
}

@article{dubessy2025quantum,
  title={Quantum gases in bubble traps},
  author={Dubessy, Romain and Perrin, H{\'e}l{\`e}ne},
  journal={AVS Quantum Science},
  volume={7},
  number={1},
  year={2025},
  publisher={AIP Publishing}
}

@article{EyinkSreenivasan2006Onsager,
  author  = {Eyink, G. L. and Sreenivasan, K. R.},
  title   = {Onsager and the Theory of Hydrodynamic Turbulence},
  journal = {Reviews of Modern Physics},
  volume  = {78},
  pages   = {87--135},
  year    = {2006},
  doi     = {10.1103/RevModPhys.78.87}
}

@article{eyink1993negative,
  title={Negative-temperature states and large-scale, long-lived vortices in two-dimensional turbulence},
  author={Eyink, G. and Spohn, H},
  journal={J. Stat. Phys.},
  volume={70},
  number={3},
  pages={833--886},
  year={1993},
  publisher={Springer}
}

@article{fetter2009rotating,
  title={Rotating trapped {Bose--Einstein} condensates},
  author={Fetter, Alexander L},
  journal={Reviews of Modern Physics},
  volume={81},
  number={2},
  pages={647--691},
  year={2009},
  publisher={APS}
}

@article{Guo2020,
  author  = {Guo, Y. and Dubessy, R. and Herve, M. d. G. d. and Kumar, A. and Badr, T. and Perrin, A. and Longchambon, L. and Perrin, H.},
  title   = {Supersonic Rotation of a Superfluid: A Long-Lived Dynamical Ring},
  journal = {Physical Review Letters},
  volume  = {124},
  pages   = {025301},
  year    = {2020},
  doi     = {10.1103/PhysRevLett.124.025301}
}

@book{galperin2019zonal,
  title     = {Zonal Jets: Phenomenology, Genesis, and Physics},
  editor    = {B. Galperin and P. L. Read},
  year      = {2019},
  publisher = {Cambridge University Press},
  isbn      = {9781107043886},
  doi       = {10.1017/9781107358225}
}

@article{gauthier2019giant,
  author  = {Gauthier, G. and Reeves, M. T. and Yu, X. and Bradley, A. S. and Baker, M. and Bell, T. A. and Rubinsztein-Dunlop, H. and Davis, M. J. and Neely, T. W.},
  title   = {Giant vortex clusters in a two-dimensional quantum fluid},
  journal = {Science},
  volume  = {364},
  pages   = {1264--1267},
  year    = {2019},
  doi     = {10.1126/science.aat5718}
}

@article{GeldhauserRomito,
  author  = {Geldhauser, C. and Romito, M.},
  title   = {The point vortex model for the Euler equation},
  journal = {Communications in Mathematical Physics},
  volume  = {361},
  number  = {2},
  pages   = {657--681},
  year    = {2018},
  doi     = {10.1007/s00220-018-3161-1}
}

@article{GuoGutierrezReyBadrPerrin2022ExpansionShellTrap,
  author  = {Guo, Y. and Gutierrez, E. D. M. and Rey, D. and Badr, T. and Perrin, A. and Longchambon, L. and Bagnato, V. S. and Perrin, H. and Dubessy, R.},
  title   = {Expansion of a quantum gas in a shell trap},
  journal = {New Journal of Physics},
  volume  = {24},
  pages   = {093040},
  year    = {2022},
  doi     = {10.1088/1367-2630/ac919f}
}

@article{Helmholtz1858,
  author  = {Helmholtz, H. v.},
  title   = {On the integrals of the hydrodynamical equations which express vortex-motion},
  journal = {Philosophical Magazine},
  volume  = {33},
  pages   = {485--512},
  year    = {1867}
}

@book{Kirchhoff1883Vorlesungen,
  author       = {Kirchhoff, G.},
  title        = {Vorlesungen über Mathematische Physik},
  volume       = {1},
  publisher    = {Teubner},
  address      = {Leipzig},
  year         = {1883},
}

@article{JoyceMontgomery1973,
  author  = {Joyce, G. and Montgomery, D.},
  title   = {Negative temperature states for a two-dimensional vortex system},
  journal = {Journal of Fluid Mechanics},
  volume  = {59},
  pages   = {209--230},
  year    = {1973}
}

@article{Kwon2014,
  author    = {Kwon, W. J. and Moon, G. and Choi, J. Y. and Seo, S. W. and Shin, Y. I.},
  title     = {Relaxation of superfluid turbulence in highly oblate {Bose-Einstein} condensates},
  journal   = {Physical Review A},
  volume    = {90},
  number    = {6},
  pages     = {063627},
  year      = {2014},
  doi       = {10.1103/PhysRevA.90.063627}
}

@article{LydonNazarenkoLaurie2022DipoleDynamicsPointVortex,
  author  = {Lydon, K. and Nazarenko, S. and Laurie, J.},
  title   = {Dipole dynamics in the point vortex model},
  journal = {Journal of Physics A: Mathematical and Theoretical},
  volume  = {55},
  number  = {38},
  pages   = {385702},
  year    = {2022},
  doi     = {10.1088/1751-8121/ac89bc}
}

@article{mattingly2002ergodicity,
  title={Ergodicity for {SDE}s and approximations: locally {L}ipschitz vector fields and degenerate noise},
  author={Mattingly, J. and Stuart, A. and Higham, D.},
  journal={Stochastic processes and their applications},
  volume={101},
  number={2},
  pages={185--232},
  year={2002},
  publisher={Elsevier}
}

@article{MerlotiDubessyLongchambonPerrin20132DQuantumGasMagnetTrap,
  author  = {Merloti, K. and Dubessy, R. and Longchambon, L. and Perrin, A. and Pottie, P. E. and Lorent, V. and Perrin, H.},
  title   = {A two-dimensional quantum gas in a magnetic trap},
  journal = {New Journal of Physics},
  volume  = {15},
  pages   = {033007},
  year    = {2013},
  doi     = {10.1088/1367-2630/15/3/033007}
}

@article{Miller1990,
  author  = {Miller, J.},
  title   = {Statistical mechanics of {Euler} equations in two dimensions},
  journal = {Physical Review Letters},
  volume  = {65},
  number  = {17},
  pages   = {2137--2140},
  year    = {1990},
  doi     = {10.1103/PhysRevLett.65.2137}
}

@article{MaestriniSalman2019,
  author  = {Maestrini, D. and Salman, H.},
  title   = {Entropy of Negative Temperature States for a Point Vortex Gas},
  journal = {Journal of Statistical Physics},
  volume  = {176},
  pages   = {981--1008},
  year    = {2019},
  doi     = {10.1007/s10955-019-02329-w}
}

@article{Nazarenko2006,
  author    = {Nazarenko, S. and Onorato, M.},
  title     = {Freely Decaying Turbulence and {Bose--Einstein} Condensation in {Gross--Pitaevskii} Model},
  journal   = {Journal of Low Temperature Physics},
  volume    = {146},
  number    = {1--2},
  year      = {2007},
  doi       = {10.1007/s10909-006-9230-4}
}

@article{NazarenkoOnorato2006WaveTurbulenceBEC,
  author  = {Nazarenko, S. V. and Onorato, M.},
  title   = {Wave turbulence and vortices in {Bose--Einstein} condensation},
  journal = {Physica D},
  volume  = {219},
  pages   = {1--12},
  year    = {2006},
  doi     = {10.1016/j.physd.2006.05.007}
}

@book{Newton2001,
  author    = {Newton, P. K.},
  title     = {The {N-}Vortex Problem: Analytical Techniques},
  publisher = {Springer},
  address   = {New York},
  year      = {2001},
  isbn      = {978-0387952260}
}

@article{Onsager1949,
  author  = {Onsager, L.},
  title   = {Statistical hydrodynamics},
  journal = {Il Nuovo Cimento},
  volume  = {6},
  pages   = {279--287},
  year    = {1949},
  doi     = {10.1007/BF02780991}
}

@article{PottersVaillantBouchet2013,
  author  = {Potters, M. and Vaillant, T. and Bouchet, F.},
  title   = {Sampling microcanonical measures of the {2D} {E}uler equations through {C}reutz’s algorithm: a phase transition from disorder to order when energy is increased},
  journal = {Journal of Statistical Mechanics: Theory and Experiment},
  volume  = {2013},
  pages   = {P02017},
  year    = {2013},
  doi     = {10.1088/1742-5468/2013/02/P02017}
}

@article{patil2021chiral,
  author  = {Patil, V. P. and Dunkel, J.},
  title   = {Chiral Edge Modes in {H}elmholtz-{O}nsager Vortex Systems},
  journal = {Physical Review Fluids},
  volume  = {6},
  pages   = {064702},
  year    = {2021},
  doi     = {10.1103/PhysRevFluids.6.064702}
}

@article{pointin1976statistical,
  title={Statistical mechanics of two-dimensional vortices in a bounded container},
  author={Pointin, Y. B. and Lundgren, T.},
  journal={The Physics of Fluids},
  volume={19},
  number={10},
  pages={1459--1470},
  year={1976},
  publisher={AIP Publishing}
}

@article{reeves2013onsager,
  author  = {Reeves, M. T. and Billam, T. P. and Anderson, B. P. and Bradley, A. S.},
  title   = {{O}nsager--{K}raichnan Condensation in Decaying Two-Dimensional Quantum Turbulence},
  journal = {Physical Review Letters},
  volume  = {110},
  pages   = {104501},
  year    = {2013},
  doi     = {10.1103/PhysRevLett.110.104501}
}

@article{ReevesGoddardLeeGauthier2022TurbulentRelaxation,
  author  = {Reeves, M. T. and Goddard-Lee, K. and Gauthier, G. and Stockdale, O. R. and Salman, H. and Edmonds, T. and Yu, X. and Bradley, A. S. and Baker, M. and Rubinsztein-Dunlop, H. and Davis, M. J. and Neely, T. W.},
  title   = {Turbulent Relaxation to Equilibrium in a Two-Dimensional Quantum Vortex Gas},
  journal = {Physical Review X},
  volume  = {12},
  pages   = {011031},
  year    = {2022},
  doi     = {10.1103/PhysRevX.12.011031}
}

@article{ReevesBillamYuBradley2017EnstrophyCascade,
  author  = {Reeves, M. T. and Billam, T. P. and Yu, X. and Bradley, A. S.},
  title   = {Enstrophy Cascade in Decaying Two-Dimensional Quantum Turbulence},
  journal = {Physical Review Letters},
  volume  = {119},
  pages   = {184502},
  year    = {2017},
  doi     = {10.1103/PhysRevLett.119.184502}
}

@article{Robert1990,
  author  = {Robert, R.},
  title   = {États d'équilibre statistique pour l'écoulement bidimensionnel d'un fluide parfait},
  journal = {Comptes Rendus de l'Académie des Sciences de Paris, Série I},
  volume  = {311},
  pages   = {575--578},
  year    = {1990}
}

@article{Robert1991,
  author  = {Robert, R.},
  title   = {A maximum-entropy principle for two-dimensional perfect fluid dynamics},
  journal = {Journal of Statistical Physics},
  volume  = {65},
  number  = {3-4},
  pages   = {531--553},
  year    = {1991},
  doi     = {10.1007/BF01053747}
}

@article{Robert2000,
  author  = {Robert, R.},
  title   = {On the statistical mechanics of {2D} {E}uler equation},
  journal = {Communications in Mathematical Physics},
  volume  = {212},
  number  = {1},
  pages   = {245--256},
  year    = {2000},
  doi     = {10.1007/s002200000225}
}

@article{RobertSommeria1991,
  author  = {Robert, R. and Sommeria, J.},
  title   = {Statistical equilibrium states for two-dimensional flows},
  journal = {Journal of Fluid Mechanics},
  volume  = {229},
  pages   = {291--310},
  year    = {1991},
  doi     = {10.1017/S0022112091003039}
}

@article{simula2014emergence,
  title   = {Emergence of Order from Turbulence in an Isolated Planar Superfluid},
  author  = {Simula, T. P. and Davis, M. J. and Helmerson, K.},
  journal = {Physical Review Letters},
  volume  = {113},
  number  = {16},
  pages   = {165302},
  year    = {2014},
  doi     = {10.1103/PhysRevLett.113.165302}
}

@article{Skipp2023HamiltonianPointVortex,
  author  = {Skipp, J. and Laurie, J. and Nazarenko, S.},
  title   = {Hamiltonian derivation of the point vortex model from the two-dimensional nonlinear {S}chrödinger equation},
  journal = {Physical Review E},
  volume  = {107},
  pages   = {025107},
  year    = {2023},
  doi     = {10.1103/PhysRevE.107.025107}
}

@article{SmithONeil1990,
  author    = {Smith, R. A. and O'Neil, T. M.},
  title     = {Nonaxisymmetric thermal equilibria of a cylindrically bounded guiding‐center plasma or discrete vortex system},
  journal   = {Physics of Fluids B},
  volume    = {2},
  number    = {12},
  pages     = {2961--2975},
  year      = {1990},
  doi       = {10.1063/1.859362}
}

@article{Seo2017,
  author  = {Seo, S. W. and Ko, B. and Kim, J. H. and Shin, Y.},
  title   = {Observation of vortex-antivortex pairing in decaying {2D} turbulence of a superfluid gas},
  journal = {Scientific Reports},
  volume  = {7},
  pages   = {4587},
  year    = {2017},
  doi     = {10.1038/s41598-017-04122-9}
}

@article{StaggAllenParkerBarenghi2015QTTBEC,
  author  = {Stagg, G. W. and Allen, A. J. and Parker, N. G. and Barenghi, C. F.},
  title   = {Generation and decay of two-dimensional quantum turbulence in a trapped {B}ose--{E}instein condensate},
  journal = {Physical Review A},
  volume  = {91},
  pages   = {013612},
  year    = {2015},
  doi     = {10.1103/PhysRevA.91.013612}
}

@article{turkington1999statistical,
  title={Statistical equilibrium measures and coherent states in two-dimensional turbulence},
  author={Turkington, B.},
  journal={Communications on Pure and Applied Mathematics},
  volume={52},
  number={7},
  pages={781--809},
  year={1999},
  publisher={Wiley Online Library}
}

@article{Wiegmann2013HydrodynamicsEulerFQHE,
  author       = {Wiegmann, P. B.},
  title        = {Hydrodynamics of Euler incompressible fluid and the fractional quantum Hall effect},
  journal      = {Physical Review B},
  volume       = {88},
  number       = {24},
  pages        = {241305(R)},
  year         = {2013},
  doi          = {10.1103/PhysRevB.88.241305},
}

@article{yatsuyanagi2005dynamics,
  author  = {Yatsuyanagi, Y. and Yasuhito, K. K. and Tomita, H. and Sano, M. M. and Yoshida, T. and Ebisuzaki, T.},
  title   = {Dynamics of Two-Sign Point Vortices in Positive and Negative Temperature States},
  journal = {Physical Review Letters},
  volume  = {94},
  pages   = {054502},
  year    = {2005},
  doi     = {10.1103/PhysRevLett.94.054502}
}

@article{YuBillamNianReevesBradley2016VortexClustering,
  author  = {Yu, X. and Billam, T. P. and Nian, J. and Reeves, M. T. and Bradley, A. S.},
  title   = {Theory of the vortex-clustering transition in a confined two-dimensional quantum fluid},
  journal = {Physical Review A},
  volume  = {94},
  pages   = {023602},
  year    = {2016},
  doi     = {10.1103/PhysRevA.94.023602}
}

@article{Carnevale1991Evolution,
  author  = {Carnevale, G. F. and McWilliams, J. C. and
             Pomeau, Y. and Weiss, J. B. and Young, W. R.},
  title   = {Evolution of Vortex Statistics in Two-Dimensional Turbulence},
  journal = {Physical Review Letters},
  volume  = {66},
  number  = {21},
  pages   = {2735--2737},
  year    = {1991},
  doi     = {10.1103/PhysRevLett.66.2735}
}

\appendix
\section{Microcanonical Monte-Carlo sampling}
\label{sec:MC}

The microcanonical equilibria described in \S\ref{sec:onsager} can be sampled using MC simulations, relying on a Markov chain variant of the Creutz algorithm~\cite{creutz1983microcanonical, SmithONeil1990, PottersVaillantBouchet2013, YuBillamNianReevesBradley2016VortexClustering}. 
To sample a target energy level $E$, the algorithm performs a random walk in the space of vortex configurations with fixed imbalance and momentum. Rather than sampling the energy shell directly, it samples an extended ensemble, which becomes equivalent in the thermodynamic limit and whose prescription depends on the sign of the temperature. 

For positive-temperature states, the target energy is used as an upper bound and the proposed update $\mathcal C \to \mathcal C'$ is accepted whenever $\mathcal{H}(\mathcal C') \leq E$. The resulting Markov chain samples the extended phase space $\Omega_N^<(E)$, which in the large-$N$ limit is dominated by configurations that lie exponentially close to the energy shell $\mathcal H=E$.  

From the entropy estimate $\Omega_N(N^2(e - \delta) )\simeq e^{NS(e)-N\beta \delta}$ in Eq.~\eqref{fullS}, it follows that  
after equilibration, the energy deficit $\delta=(E-\mathcal H)/N^2$ is exponentially distributed with rate $N\beta$ -- providing an estimate for the temperature $T=\beta^{-1}$.

Similarly, negative-temperature states are sampled by imposing the target energy level as a lower bound. In this case, the positive excess energy  $\delta=(\mathcal H-E)/N^2$ is exponentially distributed with rate $N|\beta|$.  The (suitably rescaled) negative temperature is then identified as 
$T=-1/|\beta|$; see Fig.~\ref{fig:MC}.

To sample configurations with a prescribed momentum, each proposed update $\mathcal C \to \mathcal C'$ selects two vortices at random with positions $\mathbf{r}_i$ and $\mathbf{r}_j$, and reinjects them randomly within the disk at positions $\mathbf{r}'_i$, $\mathbf{r}'_j$ such that $\Gamma_i |\mathbf{r}_i|^2+ \Gamma_j |\mathbf{r}_j|^2 =  \Gamma_i |\mathbf{r}'_i|^2+ \Gamma_j |\mathbf{r}'_j|^2$. Reinjections that take vortices outside the domain are rejected. We define one MC step as $N$ such elementary moves, while averages are typically computed over several thousands of MC steps; see Fig.~\ref{fig:MC}.

\begin{figure}[htp!]
\includegraphics[scale=0.46]{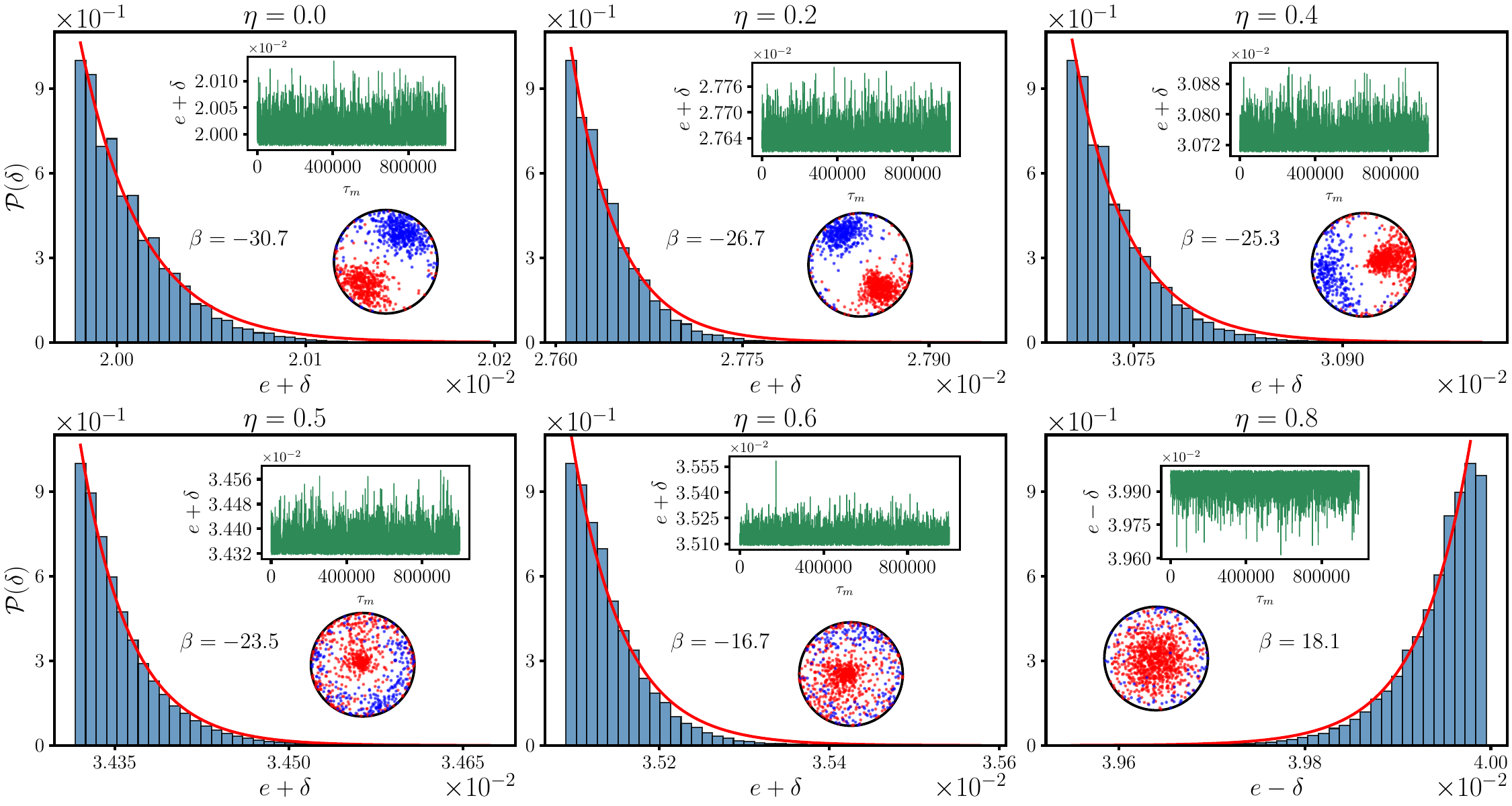}
\caption{Energy distribution obtained through MC sampling of the final state of the runs in Table~\ref{table:EM}. Insets show the MC time series. The red curves denote the exponential distributions $N|\beta| e^{-N |\beta| |\delta| }$ with $\delta$ the energy deficit or excess -- see main text. The panels corresponding to $\eta \le 0.6$ show negative-temperature sampling. The panel at $\eta=0.8$ shows a positive-temperature sampling.}
\label{fig:MC}
\end{figure}

In both cases, the MC algorithm requires an admissible initial configuration with the prescribed momentum and admissible energy.
To obtain the prescribed momentum,  we start from a random vortex configuration and repeatedly perform elementary moves where only one vortex is randomly displaced. The proposed moves are accepted if they change the momentum in the desired direction until the target momentum is reached. Once the prescribed momentum has been obtained, we perform a sequence of MC moves $\mathcal{C}\to \mathcal{C}'$, where transitions are accepted only if they increase (resp. decrease) energy, until the target energy is crossed. The resulting configuration serves as the initial state for the subsequent MC sampling.

\section{MF equilibria in the strong mixing limit}
\label{sec:linear}
We here give the technical details necessary to compute the bifurcation diagram of  \S~\ref{sec:mfeq}.
\subsection{Bridging formulae}
Recall that the linearisation of the $n^\pm$ at small $x=\beta \psi + \alpha r^2$ yields
\beq
	n_*^{\pm}(\mathbf{r}) = \left(1\mp x_*\right) \exp\left(- \gamma^{\pm}  \right),\quad x_*= \beta  \psi_*(\mathbf{r}) +  \alpha r^2 
	\label{eq:npm_lim}
\eeq
along with the linearized PB
\beq
-\Delta\psi_*=C-B\psi_*  -Ar^2 ,\quad D:=e^{-\gamma^+}+e^{-\gamma^-},\quad  C:=e^{-\gamma^+}-e^{-\gamma^-} ,\quad B=D\beta,\quad A=D\alpha.
\eeq
Integrating Eq.~\eqref{eq:npm_lim} over $\mD$, we  compute
\beq
	\dfrac{N^{\pm}}{N} = \left(\pi\mp \dfrac{X}{D}  \right) \exp\left(- \gamma^{\pm}  \right), \quad  X:= D\scal{1,x_*}=B  \scal{1,\psi_*}+A  \dfrac{\pi}{2};
\eeq
This implies 
\beq
1=\pi D- XC/D,\quad \eta =\pi C- X
\label{eq:N}
\eeq
and reversely
\beq
 D=\dfrac{1}{2\pi} \left(1 + \sqrt{1+{4 X (X+\eta )}}\right),\quad  C= \dfrac{1}{\pi}\left(X+\eta \right)
\label{eq:DC}
\eeq
To retrieve $\alpha, \beta,\gamma^\pm$ from $A,B,C,D,\psi_*$ we use the bridge formulas
\beq
	\gamma^+ = -\ln \dfrac{C+D}{2},	\quad \gamma^- = -\ln \dfrac{D-C}{2},\quad \beta = B/D,\quad \alpha=A/D.
\label{eq:bridge}
\eeq

\subsection{Solving the linearized Poisson-Boltzmann (PB)}
\label{ssec:linPB}
Decomposing the linearized PB into the Fourier-Bessel modes yield
\beq
\left(\lambda_{lm}^2+B\right)p_{lm}=C\scal{1,\phi_{lm}} -A\scal{r^2,\phi_{lm}},
\eeq
where we  explicitly compute
\beq
\scal{1,\phi_{lm}}
=
\delta_{m0}\,
\frac{2\sqrt{\pi}\,(-1)^{l+1}}{j_l} = a_l^{(0)}\delta_{m0}
\quad
\scal{r^2,\phi_{lm}}
=
\delta_{m0}\,
2\sqrt{\pi}\,(-1)^{l+1}
\frac{j_l^2-4}{j_l^3}=  a_l^{(2)}\delta_{m0},
\qquad
j_l \equiv j_l^{(0)}.
\eeq
The nature  of the solution depends on the value of $B$. \\

 \paragraph{Continuum branch:} When $B\neq-\lambda^2_{lm} $ (for any $l,m$),  the linearized PB operator and the angular and imbalance constraints uniquely determine $A$ and $C$. This yields a continuum solution of the form
\beq
	\psi_c(B,\eta,\mu) =\sum_{l\ge 1} p_{l0}\phi_{l0}, \quad  p_{l0}= \dfrac{C a_l^{(0)} - A a_l^{(2)}  }{\lambda_{l0}^2+B}
\label{eq:conti}
\eeq
The coefficients $C$ and $A$ can be expressed as a function of $\mu,\eta,B$ as
\beq
\begin{split}
& A(B,\eta,\mu)=
\frac{f_0 \mu - g\eta}
     {g^2-f_0f_2},\qquad
C(B,\eta,\mu)=
\frac{ g\mu -f_2\eta}
     {g^2-f_0f_2}\\
&  g(B):=\sum_{l\ge1}\lambda_{l0}^2 \dfrac{a_l^{(0)}a_l^{(2)} }{\lambda_{l0}^2+B},\quad f_0(B):=\sum_{l\ge1}\lambda_{l0}^2\dfrac{{(a_l^{(0)}})^2 }{\lambda_{l0}^2+B},\quad f_2(B)=\sum_{l\ge1}\lambda_{l0}^2\dfrac{ (a_l^{(2)})^2  }{\lambda_{l0}^2+B}
\end{split}
\label{eq:AC}
\eeq
These formulae  follow from the straightforward identities
\beq
	\mu=C g -Af_2,\quad
	\eta= Cf_0-Ag,
\label{eq:mueta}
\eeq
and the observation that the determinant $\Delta (B)=g^2 -f_0f_2$ is non vanishing; see \S~\ref{sec:mittlag}. In turn, the energy of the continuum state is
\beq
	\mathcal E_c (B,\eta,\mu)= \dfrac{1}{2}
 \sum_{l\ge 1}\lambda_{l0}^2    \left(\dfrac{ C a_l^{(0)} - A a_l^{(2)}  }{\lambda_{l0}^2+B}\right)^2.
\eeq
Note that
\beq
\begin{split}
	& 2 \mathcal E_c (B,\eta,\mu)=C^2 \phi_0(B)+A^2 \phi_2(B)-2AC \phi_{02}(B),\\
\quad
& \phi_0(B)=\sum_{l\ge1} \dfrac{\lambda_{l0}^2 (a_l^{(0)})^2  }{\left(\lambda_{l0}^2+B\right)^2},\qquad
\phi_2(B)=\sum_{l\ge1} \dfrac{\lambda_{l0}^2 (a_l^{(2)})^2  }{\left(\lambda_{l0}^2+B\right)^2},\qquad
\phi_{02}(B)=\sum_{l\ge1} \dfrac{\lambda_{l0}^2 (a_l^{(0)})  (a_l^{(2)})  }{\left(\lambda_{l0}^2+B\right)^2}
\end{split}
\eeq
We  show in \S~\ref{sec:mittlag},  that  the infinite sums  on the $l$ index  can be simplified using the Mittag-Leffler expansions.
The value of the energy selects admissible $B$'s.\\

\paragraph{Non-axisymmetric resonance:} When $B$ coincides with a non-axisymmetric eigenvalue, $B=-\lambda^2_{lm} , m\ge 1, m\neq 3$:
 the corresponding null mode can be added, giving a mixed solution.
In this case
\beq
	\psi = \psi_c(B=-\lambda_{l_0m_0}^2,\eta,\mu) + 2 \Re( \chi \phi_{l_0m_0}) ,\qquad \chi \in \mathbb C.
\eeq

The $\chi$ term does not contribute to the circulation or momentum. Its modulus is determined by  the energy as
\beq
		\mathcal E_m (\chi,B)=
\lambda^2_{l_0 m_0}|\chi|^2+ \mathcal E_c(-\lambda_{l_0m_0}^2,\eta,\mu),
\eeq
and its phase factor is arbitrary reflecting the possibility to add a constant polar angle to the PB solution.
\paragraph{Axisymmetric resonance:} 
When $B$ coincides with an axisymmetric eigenvalue, say   $B=-\lambda^2_{0l}$, 
the value of $p_{0l}$ is seemingly undefined but this apparent singularity is removable at fixed $\eta$ and $\mu$. 
Writing $B=-\lambda^2_{0l}+ \epsilon$, one finds
$Ca_{l}^{(0)}-Aa_{l}^{(2)}=O(\epsilon),$ and hence
$p_{0l}(-\lambda^2_{0l}+\epsilon,\eta,\mu)=O(1)$, meaning that the resonance  is part of the continuum and does not generate an independent  mixed branch.\\

\paragraph{Degenerate resonance:} When $B=-\lambda_{l3}^2=-\left(j^{(3)}_l\right)^2$ then  $\Delta (B)=0$. In this case, 
 $\mu=\eta/2$ and $g=f_0/2$, implying that the relations \eqref{eq:mueta}  are not independent, and reduce to the single equation
\beq
\eta =\left(C-A/2 \right)f_0(-\lambda_{l3}^2)
\eeq
 The solution of the PB has the two-parameter form

\beq
	\psi = \tilde \psi_c(C) + 2 \Re({\chi \phi_{l_03}})\quad  \qquad \chi\in \mathbb C, C \in \mathbb R.
\eeq

involving the degenerate continuum background
\beq
	\tilde \psi_c(C) =\sum_{l\ge 1} \tilde p_{l0}\phi_{l0}, \quad  \tilde{p}_{l0}= \dfrac{C a_l^{(0)} - \tilde A a_l^{(2)}  }{\lambda_{l0}^2-\lambda_{l3}^2},\quad 
\tilde A=2C-2\eta/f_0(-\lambda_{l3}^2)
\eeq

The parameters  $C,\chi$ are constrained by the total  energy.

\subsection{Maximum entropy solution}

Summarizing, away from the degenerate case $\mu = \eta/2$, the solutions belong either to an axisymmetric continuum branch or to mixed branches occurring at non-axisymmetric resonances, as prescribed by Eq.~\eqref{eq:genpsi}.
When several branches coexist at the same values of $(e,\eta,\mu)$, the optimization problem  selects the one which maximizes the entropy in Eq.~\eqref{eq:Sstar}.
The entropy requires the bridging relations \eqref{eq:bridge} to compute the Lagrange multipliers $\alpha, \beta,\gamma^\pm$,
along with the observation that the $X$-factor in  Eq.~\eqref{eq:DC} relies solely on the continuum component as 
\beq
	X= B C F_0(B) +A \left(\dfrac{\pi}{2}-BG(B)\right),\quad \text{for }   G(B):=\sum_{l\ge1}\dfrac{a_l^{(0)}a_l^{(2)} }{\lambda_{l0}^2+B},\quad F_0(B):=\sum_{l\ge1}\dfrac{{(a_l^{(0)}})^2 }{\lambda_{l0}^2+B}
\label{eq:X}
\eeq

\subsection{Mittag-Leffler expansions}
\label{sec:mittlag}
The various functional sums  appearing in \S \ref{ssec:linPB}
admit explicit expressions  in terms of Bessel functions, using so-called Mittag-Leffler expansions.  Those expressions prove useful for the numerical construction of the phase diagram, and we therefore list them below. For  $B\le 0$,  we set $z=\sqrt{-B}$. For  $B\ge 0$,  we set $z=i\sqrt{|B|}$.
\begin{enumerate} 

	\item $f_0(B) = \sum_{l\ge1}\lambda_{l0}^2 \dfrac{{(a_l^{(0)}})^2 }{\lambda_{l0}^2+B} =\dfrac{2\pi J_1(z)}{zJ_0(z)}$
	\item $f_2(B) = \sum_{l\ge1}\lambda_{l0}^2 \dfrac{{(a_l^{(2)}})^2 }{\lambda_{l0}^2+B} = 
\dfrac{(B+4)^2 f_0(B) -2\pi (3B+8)}{B^2}$
	\item $g(B) = \sum_{l\ge1}\lambda_{l0}^2 \dfrac{a_l^{(0)}a_l^{(2)}} {\lambda_{l0}^2+B} = \dfrac{B+4}{B} f_0(B)-\dfrac{4\pi}{B}$

	 \item   $F_0(B):=\sum_{l\ge1}\dfrac{{(a_l^{(0)}})^2 }{\lambda_{l0}^2+B}$: \quad $BF_0+f_0=\pi$
	 \item   $F_2(B):=\sum_{l\ge1}\dfrac{{(a_l^{(2)}})^2 }{\lambda_{l0}^2+B}$: \quad $BF_2+f_2= \dfrac{\pi}{3}$
	\item   $G(B):=\sum_{l\ge1}\dfrac{a_l^{(0)}a_l^{(2)} }{\lambda_{l0}^2+B},$	\qquad $ BG + g =\dfrac{\pi}{2}$

	\item $\phi_0(B)=\sum_{l\ge1} \dfrac{\lambda_{l0}^2 (a_l^{(0)})^2  }{\left(\lambda_{l0}^2+B\right)^2} = \dfrac{\pi}{z^2}\left(\dfrac{J_1(z)^2}{J_0(z)^2}-2\dfrac{J_1(z)}{zJ_0(z)}+1\right)$

	\item $\phi_2(B)=\sum_{l\ge1} \dfrac{\lambda_{l0}^2 (a_l^{(2)})^2  }{\left(\lambda_{l0}^2+B\right)^2}=\dfrac{(B+4)^2}{B^2}\phi_0(B)
+\dfrac{8(B+4)}{B^3}f_0(B)
-\dfrac{2\pi(3B+16)}{B^3}$.

	\item 
$\phi_{02}(B)=\sum_{l\ge1} \dfrac{\lambda_{l0}^2 (a_l^{(0)})  (a_l^{(2)})  }{\left(\lambda_{l0}^2+B\right)^2}=\dfrac{B+4}{B}\phi_0(B)
+\dfrac{4}{B^2}f_0(B)-\dfrac{4\pi}{B^2}$

	\item $a_l^{(0)} = \scal{1,\phi_{l0}} = \frac{2\sqrt{\pi}\,(-1)^{l+1}}{j_l}$ ,\qquad $a_l^{(2)} =\langle r^2,\phi_{l0}\rangle =2\sqrt{\pi}\,(-1)^{l+1}\frac{j_l^2-4}{j_l^3}$,\qquad $j_l \equiv j_l^{(0)}$
	
	\item $\Delta(B)=g^2-f_0f_2=-\dfrac{4\pi^2}{z^{3}}
\dfrac{J_3\!\left(z\right)}
     {J_0\!\left(z\right)}$.  
\end{enumerate}

\end{document}